\documentclass[aps,twocolumn,floatfix,superscriptaddress,showpacs,nofootinbib]{revtex4}
\usepackage{graphicx}
\usepackage{amsmath}
\usepackage{amssymb}
\usepackage{bm}

\begin{document}


\title{A Computational Study of Rotating Spiral Waves\\
and Spatio-Temporal Transient Chaos\\ in a Deterministic
Three-Level Active System}
\author{S.~D.~Makovetskiy}%
\email{sdmakovetskiy@ukr.net} \affiliation{Department of Computer
Sciences, Kharkiv National University of Radio Electronics,\\ 14,
Lenin avenue, Kharkiv, 61166, Ukraine}

\author{D.~N.~Makovetskii}%
\email{makov@ire.kharkov.ua} \affiliation{Department of Acoustic
and Electromagnetic Spectroscopy,\\ Institute of Radio-Physics and
Electronics of National Academy of Sciences,\\ 12, Academician
Proskura street, Kharkiv, 61085, Ukraine}

\date{March 10, 2005}

\begin{abstract}

Spatio-temporal dynamics of a deterministic three-level cellular
automaton (TLCA) of Zykov-Mikhailov type (Sov. Phys. -- Doklady,
1986, Vol.31, No.1, P.51) is studied numerically. Evolution of
spatial structures is investigated both for the original
Zykov-Mikhailov model (which is applicable to e.~g.
 Belousov-Zhabotinskii chemical reactions) and for proposed
by us TLCA, which is a generalization of Zykov-Mikhailov model for
the case of two-channel diffusion. Such the TLCA is a minimal model
for an excitable medium of microwave phonon laser, called phaser
(D. N. Makovetskii, Tech. Phys., 2004, Vol.49, No.2, P.224;
cond-mat/0402640). The most interesting observed manifestations of
TLCA dynamics are as follows: (a)~spatio-temporal transient chaos
in form of highly bottlenecked collective evolution of excitations
by rotating spiral waves (RSWs) with variable topological charges;
(b)~competition of left-handed and right-handed RSWs with
unexpected features, including self-induced alteration of integral
effective topological charge; (c)~transient chimera states, i.~e.
coexistence of regular and chaotic domains in TLCA patterns;
(d)~branching of TLCA states with different symmetry which may lead
to full restoring of symmetry of imperfect starting pattern.
Phenomena (a) and (c) are directly related to phaser dynamics
features observed earlier in real experiments at liquid helium
temperatures on corundum crystals doped by iron-group ions.

\vspace{6pt}

\noindent
\begin{footnotesize}ACM classes: F.1.1~Models of
Computation (Automata), I.6~Simulation and Modeling
(I.6.3~Applications), J.2~Physical Sciences and Engineering
(Chemistry, Physics)
\end{footnotesize}

\end{abstract}


\maketitle

\section{INTRODUCTION}\label{se:i}

During last years vortices and, in particular, rotating spiral
waves (RSWs) are widely studied in excitable media
\cite{Elkin-etc,Field-1985,Ramos-2001}, lasers
\cite{Staliunas-1993,Weiss-1993,Staliunas-1995,Weiss-1999,Weiss-2004},
Bose-Einstein condensates \cite{Ghosh-2004}, saturable nonlinear
media \cite{YangJ-2002}, ferromagnetics
\cite{Kandaurova-2002,Mol-2002}, photorefractive optical lattices
\cite{YangJ-2003}, open microwave billiards \cite{Kim-2003},
three-level spatial Lotka-Volterra's type models
\cite{Szabo-2002,Provata-2003} etc.

Evolution of such complex spatio-temporal structures, as patterns
of RSWs, is now a subject of extensive computer modeling,
especially for intrinsically unstable systems possessing
hypersensitivity to initial and/or border conditions, to
imperfections of active medium etc.

Choice of appropriate model is the crucial point for computer
studying of these complex systems. Even fully deterministic model
of continuous unstable system may demonstrate unpredictable
behaviour due to exponential increasing of small deviation of
initial conditions from their precise values \cite{Lorenz-etc}.
This unpredictability cannot be eliminated because of finite
precision of initial data, available for a digital computer, even
if no round-off errors exist in certain ``perfect'' digital
machine. Generally, instead of computing of the evolution of a
continuous unstable system (in the framework of given model), a
digital machine may compute an evolution of some composite system,
which includes the computer itself.

Another serious trouble is the large dimension problem, which is
typical for modeling of many-particle systems. Using of appropriate
averaging is a good solution for systems near equilibrium, but for
strongly nonequlibrium systems it often is not so.

A possible way to bypass these troubles consists in using of
partially or fully discrete analogs of continuous models
\cite{Bogach-1973,Bogach-1979,Fisch-1991,Greenberg-etc,Hordijk-1999,Kaneko-1986,Loskutov-1990,Reshodko-etc,Toffoli-etc,Toffoli-1987,Vanag-1999,Zykov-1986}.
Construction of such models is very nontrivial task, but adequate
discrete model is much more suitable for numerical investigation of
unstable and/or multiparticle systems, than, say, a model based on
partial differential equations.

In this work, we study evolution of RSWs and some other
spatio-temporal structures in discrete active (excitable) system
consisting of locally interacting three-level particles (units)
with finite times of relaxation. This active system is described by
Zykov-Mikhailov (ZM) model, which was primarily introduced in 1986
for chemical excitable systems \cite{Zykov-1986} (see also
\cite{Loskutov-1990}).

A modification of ZM model is proposed by us to adapt it for
emulation of some aspects of dynamics of microwave phonon lasers
\cite{DNM-Diss-1983,Peterson-1969,Tucker-etc} (called also phasers
\cite{Phaser-ruby-early,Nickel-early,DNM-Diss-1983}) which
demonstrate strong instabilities, cooperative behaviour,
deterministic chaos and other manifestations of complexity
\cite{DNM-Diss-1983,Tashkent-1991-etc,TPL-2001,arx-selforg-2003,TP-2004,arx04a-destab-2004}.
The relaxational properties of active units (paramagnetic ions) in
phasers are of the same type, as in class-B lasers
\cite{Tredicce-1985}, and there are experimental confirmations
\cite{TPL-2001,arx-selforg-2003,TP-2004,arx04a-destab-2004} of
common properties of phasers and this class of lasers.

In class-B lasers the inequality $T_1 \gg T_{\text{field}} \gg T_2$
\cite{Tredicce-1985} takes place, where $T_{\text{field}}$ is the
lifetime of {\textrm{photons}}; $T_1$ and $T_2$ are respectively
the times of longitudinal and transverse relaxation of active
units. The same inequality we have for phasers, if one replace
{\textit{photons}} by {\textit{phonons}} (see
\cite{DNM-Diss-1983,TPL-2001,arx-selforg-2003,TP-2004,arx04a-destab-2004}).
From this point of view phasers are acoustic analogs of class-B
lasers. Moreover, acoustic wavelenghts in phasers is of the same
order as electromagnetic ones in optical lasers, because sound
velocity in phaser is about 5 orders less than light velocity in
laser (and phasers usually operate at frequencies $F =
F^{({\mathrm{phaser}})} \approx 10^{-5} F^{({\mathrm{laser}})}$
\cite{Phaser-ruby-early,Nickel-early,DNM-Diss-1983,Tashkent-1991-etc,TPL-2001,arx-selforg-2003,TP-2004,arx04a-destab-2004,Peterson-1969,Tucker-etc}).

In papers \cite{Weiss-1993,Staliunas-1995,Weiss-2004} class-B
lasers were treated as a kind of excitable systems, that are widely
studied in chemical kinetics, especially for the
Belousov-Zhabotinskii (BZ) reaction and other similar chemical
phenomena
\cite{Agladze-etc,Field-1985,Kapral-1995,Ramos-2001,Winfree-1972}.
Some of differences and similarities between class-B lasers and
excitable systems were examined in \cite{Staliunas-1995}:

\begin{quotation}
``The difference is that excitable systems in zero-dimensional case
(with spatially homogeneous fields) display self-sustained
oscillations of the field amplitude, whereas ... [class-B systems]
display (damped) relaxation oscillations only. In the limit%
\footnote{In \cite{Staliunas-1995}, $\gamma_\parallel \equiv
T_{\text{field}} / T_1$.}
$\gamma_\parallel \rightarrow 0$ the relaxation oscillations decay
very slowly; thus class-B lasers may be considered quasi-excitable
systems in this limit. In two spatial dimensions (unlike the
zero-dimensional case) class-B lasers display self-sustained
oscillations: They behave quite similarly to excitable systems. The
non-stationarity of the vortices \cite{Weiss-1993} is one
indication of the similarity with the excitable systems.''
(\cite{Staliunas-1995}, page 1143).
\end{quotation}

Using pointed similarity and taking into account the analogy
between phasers and class-B lasers, a phaser dynamics in
two-dimension (2D) case may be emulated by ZM model
\cite{Zykov-1986,Loskutov-1990} (with some extensions of the last,
concerning relaxational properties of active units). The ZM model
is a cellular automaton (CA) \cite{vonNeumann-1966}, i.e. discrete
mapping, which may be defined as follows. Let an active discretized
medium ${\mathfrak{M}}_{\mathrm{e}}$ has the form of rectangular
${\mathrm{2D}}$\nobreakdash-\hspace{0pt}lattice. Each cell of the
lattice contains one cellular-automaton unit (CAU) with coordinates
$(i,j)$, where $(\min(i) = 1) \wedge (\min(j) = 1)$. All the CAUs
in the ${\mathfrak{M}}_{\mathrm{e}}$ are identical, and they
interact by the same set of rules (CA is homogeneous and isotropic
in the von~Neumann sense \cite{vonNeumann-1966}). The upgrade of
state $S_{ij}^{(n)} \equiv S^{(n)}(i,j)$ of each CAU is carried out
synchronously at each step $n \leq N$ during the cellular automaton
evolution. The final step $N$ may be either predefined or it will
be searched during CA evolution as time (i.e. quantity of discrete
steps) for reaching an attractor of this CA. The conditions of
upgrade depends both on $S_{ij}^{(n-1)}$ and on $S_{i'j'}^{(n-1)}$,
where $\{i',j'\}$ belongs to certain active neigborhood of CAU at
site $(i,j)$. In order to formulate boundary conditions for a
${\mathfrak{M}}_{\mathrm{e}}$ of finite size (i.~e. when $\max(i) =
M_X, \max(j) = M_Y$), a set ${\mathfrak{M}}_{\mathrm{v}}$ of
virtual cells with coordinates $(i = 0) \vee (j = 0) \vee (i = M_X
+ 1) \vee (j = M_Y + 1)$ may be introduced.

The original ZM model has the single channel of diffusion
\cite{Zykov-1986,Loskutov-1990}. Such one-channel (1C) models are
adequate for chemical reaction-diffusion systems \cite{Tyson-1985}.
In phaser active system, the multichannel diffusion of spin
excitations is the typical case, because it proceeds via
(near)-resonant dipole-dipole ($d-d$) magnetic interactions between
paramagnetic ions \cite{DNM-Diss-1983}. For a three-level system,
which is the simplest phaser system, there are 3 possible channels
of resonant diffusion. In the case when $d-d$ interactions are
forbidden at one of three resonant frequencies of a three-level
system, the asymmetric two-channel (2C) diffusion is realized under
conditions of perfect refractority of the intermediate level ---
see Appendix \ref{ap:A}. Note that asymmetric diffusion is well
known in biology (see, e.g., the book of D.~A.~Frank-Kamenetzky
\cite{Frank-1984}). Recently a kind of asymmetric diffusion was
proposed by N.~Packard and R.~Shaw \cite{Packard-2004} for a
mechanical system.

Another important feature of phaser is extremely low level of
unavoidable (quantum) noise in its active medium because
$I_{\mathrm{spont}} \propto F^3$, where $I_{\mathrm{spont}}$ is
intensity of spontaneous emission. Accordingly,
$I_{\mathrm{spont}}^{({\mathrm{phaser}})}$ is about 15 orders lower
than $I_{\mathrm{spont}}^{({\mathrm{laser}})}$, because
$F^{({\mathrm{phaser}})}/F^{({\mathrm{laser}})} \approx 10^{-5}$.
It gives the basis to consider the phaser system as close to
deterministic, i.~e. one need not include stochastic terms
\cite{Vanag-1999} in CA rules for phaser modeling.

The paper is organized as follows. In Section \ref{se:ii} we review
deterministic CA generating vortices, giving main attention to
those of CA, in which dynamically stable RSWs are possible. Section
\ref{se:iii} is devoted to detailed description of ZM-like TLCA
model used in our computer experiments. The results of
computational study of RSWs and discussion are presented in
Sections \ref{se:iv} and \ref{se:v}. Section \ref{se:vi} contains
some concluding remarks. In Appendices \ref{ap:A} and \ref{ap:B},
the structure of energy levels and relaxation properties of the
${\mathrm{Ni^{2+}:Al_2O_3}}$ spin-system (used in real experiments
with phaser) are discussed. Phenomena of inversion states collapse
and critical slowing down for lumped (point-like) phaser models are
described in Appendix \ref{ap:C} in order to compare them with
slowing down phenomena in TLCA which is a minimal multiparticle
model of an active (excitable) medium of phaser.

\section{ROTATING SPIRAL WAVES IN CELLULAR AUTOMATA:
A SHORT REVIEW}\label{se:ii}

Probably the simplest cellular automaton which generates rotating
(but \textit{nonspiral}) objects is the Conway's Game of Life (CGL)
\cite{Berlecamp-1982}. CGL is a two-level ($L_G \in \{L_0, L_1\}$)
cellular automaton defined on square grid
(${\mathfrak{M}}_{\mathrm{e}}$ in CGL is usually infinite). The
state of each cell with coordinates $(i,j)$ at step $n$ of
evolution is described by very simple relationship $G =
\Phi_{ij}^{(n)}$. Here $\Phi_{ij}^{(n)}$ is binary phase counter:
$\Phi_{ij}^{(n)} \in \{0,1\}$ with such the rules of upgrading:
\begin{equation}
  \Phi_{ij}^{(n+1)} =
  \begin{cases}
    \Phi_{ij}^{(n)}, & \text{if $U_{ij}^{(n+1)} = 2$};\\
    1,               & \text{if $U_{ij}^{(n+1)} = 3$};\\
    0,               & \text{otherwise},
  \end{cases}
  \label{eq:01}
\end{equation}
where
\begin{equation}
  U_{ij}^{(n+1)} =\sum_{i'j'} \Phi_{i'j'}^{(n)},
  \label{eq:02}
\end{equation}
and the pair of indexes $\{i',j'\} \equiv \{i+a,j+b\}$ describes
the active neighborhood for the cell $(i,j)$. For the classical CGL
\cite{Berlecamp-1982}, this is the Moore neighborhood, i.e. octet
of the nearest cells: $\left[ (|a| \leq 1) \wedge (|b| \leq 1)
\right] \wedge (\delta_{a0}\delta_{b0} \neq 1)$, where
$\delta_{xy}$ is the Kronecker symbol.

In CGL, rotating (and simultaneously moving) structures called
gliders are dynamically stable if there are no another CGL objects
in their vicinity. Gliders are of great interest from the
mathematical point of view, especially because it is possible to
construct universal computing machine on the basis of gliders and
glider guns. On the other hand, CGL (at least in its classical
two-level form) does not describe any known real-world vortices.

Much more realistic description of vorticity, including RSWs, is
possible in framework of three-level cellular automata (TLCA) or
many-level ones. The simplest TLCA which generates rotating
\textit{spiral} objects is the single-step relaxation (SSR) model,
which was extensively investigated by L.~V.~Reshodko with
co-authors at the beginning of 1970-th
\cite{Bogach-1973,Bogach-1979,Reshodko-etc}. Later a similar CA was
independently proposed and studied by J.~Greenberg and S.~Hastings
(GH model) \cite{Greenberg-etc}.

\subsection{The single-step relaxations model}\label{subse:ii-A}

In SSR model, every CAU at each step of evolution $n$ occupies one
of the three discrete levels $L_K \in \{ L_{\mathrm{I}},
L_{\mathrm{III}}, L_{\mathrm{II}} \}$. The order of levels in curly
brackets corresponds to the order of CAU's state advancing
direction (by cycle): $L_{\mathrm{I}} \rightarrow L_{\mathrm{III}}
\rightarrow L_{\mathrm{II}} \rightarrow L_{\mathrm{I}} \rightarrow
\cdots$. Here $L_{\mathrm{I}}$ is the ground (lowest) level, which
is stable in the absence of nearest neighbors with $K \neq
{\mathrm{I}}$; $L_{\mathrm{III}}$ is the excited (upper) level, and
the $L_{\mathrm{II}}$ is the refractory (intermediate) level. This
``arrangement'' of levels may be described by apparently formal
(from the mathematical point of view) inequality $L_{\mathrm{III}}
> L_{\mathrm{II}} > L_{\mathrm{I}}$. But from the physical point of
view such the inequality is well defined (see Appendix \ref{ap:A}
for an example of the concrete physical system).

The essence of the SSR model is as follows
\cite{Bogach-1979,Greenberg-etc,Reshodko-etc}. Let us introduce the
phase counters $\phi_{ij}^{(n)} \in \{-1, 0, +1\}$. In the SSR
model, there is such the correspondence between $L_K$ and
$\phi_{ij}^{(n)}$:
\begin{gather}
  \bigl( L_K = L_{\mathrm{I}} \bigr) \Leftrightarrow
  \bigl( \phi_{ij}^{(n)} = 0 \bigr); \label{eq:03}\\
  \bigl( L_K = L_{\mathrm{III}} \bigr) \Leftrightarrow
  \bigl( \phi_{ij}^{(n)} = +1 \bigr); \label{eq:04}\\
  \bigl( L_K = L_{\mathrm{II}} \bigr) \Leftrightarrow
  \bigl( \phi_{ij}^{(n)} = -1 \bigr). \label{eq:05}
\end{gather}
The iterative process (which governs the whole SSR cellular
automaton evolution) describes transitions of every individual CAU
by mapping of phase counters $\phi_{ij}^{(n)}$:
\begin{equation}
  \phi_{ij}^{(n+1)} =
  a_{ij}^{(n+1)} \bigl( \phi_{ij}^{(n)} \bigr) +
  d_{ij}^{(n+1)} \bigl( \phi_{ij}^{(n)}, \phi_{i+r,j+s}^{(n)} \bigr),
    \label{eq:06}
\end{equation}
where
\begin{equation}
  a_{ij}^{(n+1)} \bigl( \phi_{ij}^{(n)} \bigl) =
  \begin{cases}
    -1, & \text{if $\phi_{ij}^{(n)} = +1$};\\
    0,  & \text{otherwise},
  \end{cases}
  \label{eq:07}
\end{equation}
and
\begin{equation}
\begin{split}
 & \qquad\qquad\qquad d_{ij}^{(n+1)} \bigl( \phi_{ij}^{(n)}, \phi_{i+r,j+s}^{(n)} \bigr) =\\
 &\phantom{a}\\
 &=
  \begin{cases}
    +1, & \text{if $\left[ \bigl( \phi_{ij}^{(n)} = 0 \bigr) \wedge \bigl( \, \exists \, \phi_{i+r,j+s}^{(n)} = +1 \bigr) \right]$};\\
    0,  & \text{otherwise}.
  \end{cases}
  \label{eq:08}
\end{split}
\end{equation}
Here the pair of indexes $\{i+r,j+s\}$ describes the von Neumann
neighborhood for the cell $(i,j)$. This neighborhood consists of
non-diagonal nearest cells only: $(|r| = 1) \oplus (|s| = 1)$ (the
symbol $\oplus$ means exclusive OR).

The SSR cellular automaton is one of the simplest 2D discrete
mappings, which models RSWs in excitable media. But the original
SSR model has insufficient resources to model real physical or
chemical excitable systems. In particular, transitions in simplest
cellular automata of the SSR type are ``instant'': the lifetimes of
excited and refractory states are precisely equal to the single
step of an iterative process.

Improving of the SSR model may be achieved by some different ways:
increasing of level numbers (``colors'') and/or extending of the
active neighborhood (see R.~Fisch, J.~Gravner, and D.~Griffeath
\cite{Fisch-1991} for details). There are also qualitatively
different models which are able to generate RSW, as the 2D
Oono-Kohmoto CA, described in the following Subsection.

\subsection{The 2D Oono-Kohmoto model}\label{subse:ii-B}

The well-known 1D Oono-Kohmoto model (Y.~Oono and M.~Kohmoto
\cite{Oono-1985}) may be generalized for the case of 2D active
medium (see paper of G.~G.~Malinetskii and M.~S.~Shakaeva
\cite{Malinetskii-1992}). Let every CAU has three levels $\{L_0,
L_1, L_M\}$, which corresponds to three discrete values of certain
global attribute $\chi$ (e.~g. concentration of a chemical agent):
$\chi_{ij}^{(n)} \in \{0, 1, {\mathcal{M}}\}$, where ${\mathcal{M}}
> 0$. The evolution of the Oono-Kohmoto 2D (OK2) cellular automaton
is described by such the discrete mapping:

\begin{equation}
  \chi_{ij}^{(n+1)} =
  \begin{cases}
    1, & \text{if $\widetilde\chi_{ij}^{(n)} \geq h_2$};\\
    0, & \text{if $h_1 \leq \widetilde\chi_{ij}^{(n)} < h_2$};\\
    {\mathcal{M}}, & \text{if $\widetilde\chi_{ij}^{(n)} < h_1 $},
  \end{cases}
  \label{eq:09}
\end{equation}
where
\begin{equation}
  \widetilde\chi_{ij}^{(n)} =
  (\alpha / {\mathcal{N}})\sum_{i'j'} \chi_{i'j'}^{(n)} + (1 -
  \alpha)\chi_{ij}^{(n)}.
  \label{eq:10}
\end{equation}

Here $h_1, h_2$ are lower and upper thresholds for
$\widetilde\chi_{ij}^{(n)}$ (e.~g. $h_1 = 1/2$; $h_2 = 3/2$), and
$\alpha$ characterizes the diffusion rate. The pair of indexes
$\{i',j'\} \equiv \{i+p,j+q\}$ describes the active neighborhoods
for the cell $(i,j)$; ${\mathcal{N}}$ is the coordination number
(quantity of cells in the active neighborhood). This may be, e.~g.,
the von~Neumann neighborhood (${\mathcal{N}}=4$), the Moore
neighborhood (${\mathcal{N}}=8$) etc.

G.~G.~Malinetskii and M.~S.~Shakaeva revealed RSWs, gliders and
other interest objects in the OK2 cellular automaton
\cite{Malinetskii-1992}. But the OK2 model (as far as the SSR one)
does not take into consideration relaxation processes. In the next
Subsection we describe some CA models with multi-step relaxation
(MSR) mechanisms.

\subsection{Multi-step relaxation models of excitable media}\label{subse:ii-C}

Introducing of finite times of relaxation (i.e. delay of interlevel
transition, which is independent of state of an active
neighborhood) leads to an important improving of cellular-automata
models of excitable media. Each of non-ground levels in an MSR
model has some virtual ``sublevels'', and relaxation of CAU may be
formally described as subsequent jumps between these ``sublevels''.

An extensive investigation of various three-level CA with MSR was
fulfilled by L.~V.~Reshodko with co-authors in early 1970-th
\cite{Bogach-1973,Bogach-1979,Letichevskii-1972,Reshodko-etc} to
model excitations in smooth muscle tissue. The starting point of
their work was three-level Wiener-Rosenblueth (WR) model
\cite{Wiener-1946}, elaborated in 1946 by N.~Wiener and
A.~Rosenblueth for continuous media. In the original WR model
\cite{Wiener-1946}, only the refractory level had finite relaxation
time. The original WR model was modified by A.~Rosenblueth in 1958
\cite{Rosenblueth-1958} with the purpose to take into consideration
finite-time relaxation of excited level. The WR model was developed
in \cite{Wiener-1946} to model RSWs in inhomogeneous media only.
But, as it was shown by O.~Selfridge \cite{Selfridge-1948} and
I.~S.~Balakhovskii \cite{Balakhovskii-1965}, such the waves are
possible in a parametrically homogeneous excitable medium too.

Another successful attempt to use of such the approach was
undertaken by T.~Toffoli and N.~Margolus (their results, to our
knowledge, were firstly published in 1987 in the book
\cite{Toffoli-1987}). About one year earlier (in 1986) V.~S.~Zykov
and A.~S.~Mikhailov had published paper \cite{Zykov-1986}, in which
a simple and very clear TLCA was proposed to model excitable media
with arbitrary relaxation times (both at excited and refractory
levels), arbitrary factor of activator accumulation (at ground
level) and arbitrary active neighborhood. In essence, the model of
Zykov and Mikhailov \cite{Zykov-1986} is a generalization of Bogach
and Reshodko (BR) model \cite{Bogach-1973,Bogach-1979}.

Birth, evolution, interaction and decay of RSWs and other
spatio-temporal structures in active (excitable) media are very
sensitive to mechanisms of diffusion. Both BR
\cite{Bogach-1973,Bogach-1979} and ZM
\cite{Zykov-1986,Loskutov-1990} models are based on the 1C
diffusion mechanism, which seems to be a good approximation for
some chemical systems (see, e.g., \cite{Tyson-1985}). Realistic
models of several physical excitable system (class-B lasers
\cite{Weiss-1993,Staliunas-1995,Weiss-2004}, phasers
\cite{DNM-Diss-1983,Tashkent-1991-etc,TPL-2001,arx-selforg-2003,TP-2004,arx04a-destab-2004}
etc.) need to take into consideration additional channels of
diffusion, as it was discussed in Section \ref{se:i}. In the next
Section we present a detailed description of the MSR model of
excitable system, based on ZM model and modified by us in order to
introduce such an additional channel in TLCA. This modified ZM
cellular automaton with 2C diffusion was used in the present paper
as the basic model for computer studying of RSWs and transient
spatio-temporal chaos in active (excitable) media.

\section{THREE-LEVEL CELLULAR AUTOMATON WITH TWO-CHANNEL DIFFUSION
(A~MODIFIED ZYKOV-MIKHAILOV MODEL)}\label{se:iii}

\subsection{States of cellular automaton units and branches of
evolution operator in TLCA}\label{subse:iii-A}

Let ${\mathfrak{M}}_{\mathrm{e}}$ is a rectangular 2D lattice
containing $M_X \times M_Y$ three-level CAUs. The upgrade of state
$S_{ij}^{(n)}$ of each CAU is (as usually) carried out
synchronously during the cellular automaton evolution. The excited
(upper) level $L_{\mathrm{III}}$ has the time of relaxation $\tau_e
\geq 1$, and the refractory (intermediate) level $L_{\mathrm{II}}$
has the time of relaxation $\tau_r \geq 1$ (as in the original ZM
model \cite{Zykov-1986} and in contrary to the SSR model, where the
equality $\tau_e = \tau_r = 1$ is embedded in). Both $\tau_e$ and
$\tau_r$ are integer numbers.

In the TLCA model with 2C diffusion, the first channel of diffusion
accelerates the transitions $L_{\mathrm{I}} \rightarrow
L_{\mathrm{III}}$ for a given CAU, and the second channel of
diffusion (which is absent in the ZM model) accelerates the
transitions $L_{\mathrm{III}} \rightarrow L_{\mathrm{II}}$. The
complete description of CAU's states $S_{ij}^{(n)}$ in such the
TLCA model includes one type of global attributes (the phase
counters $\varphi_{ij}^{(n)}$) and two types of partial attributes
$u_{ij}^{(n)}$ and $z_{ij}^{(n)}$ for each individual CAU in
${\mathfrak{M}}_{\mathrm{e}}$. Full description of all CAU's
possible states is as follows:
\begin{equation}
  S_{ij}^{(n)} \bigl( L_{\mathrm{I}} \bigr) =
  \bigl( \varphi_{ij}^{(n)}, u_{ij}^{(n)} \bigr);
  \label{eq:11}
\end{equation}
\begin{equation}
  S_{ij}^{(n)} \bigl( L_{\mathrm{III}} \bigr) =
  \begin{cases}
    \bigl( \varphi_{ij}^{(n)} \bigr),               & \text{if $n = 0$},\\
    \bigl( \varphi_{ij}^{(n)}, z_{ij}^{(n)} \bigr), & \text{if $n \neq 0$};
  \end{cases}
  \label{eq:12}
\end{equation}
\begin{equation}
  S_{ij}^{(n)} \bigl( L_{\mathrm{II}} \bigr) =
  \bigl( \varphi_{ij}^{(n)} \bigr).
  \label{eq:13}
\end{equation}

In the framework of the TLCA model, the phase counters lie in the
interval $\varphi_{ij}^{(n)} \in [0, \tau_e + \tau_r]$. The
following correspondences between $\varphi_{ij}^{(n)}$ and $L_K$
take place (by definition) for all the CAUs in
${\mathfrak{M}}_{\mathrm{e}}$ at all steps $n$ of evolution:
\begin{gather}
  \bigl( L_K = L_{\mathrm{I}} \bigr) \Leftrightarrow
  \bigl( \varphi_{ij}^{(n)} = 0 \bigr); \label{eq:14}\\
  \bigl( L_K = L_{\mathrm{III}} \bigr) \Leftrightarrow
  \bigl( 0 < \varphi_{ij}^{(n)} \leq \tau_e \bigr); \label{eq:15}\\
  \bigl( L_K = L_{\mathrm{II}} \bigr) \Leftrightarrow
  \bigl( \tau_e < \varphi_{ij}^{(n)} \leq \tau_e + \tau_r \bigr).
  \label{eq:16}
\end{gather}

These correspondences are of key significance for all MSR models
(BR model \cite{Bogach-1973,Bogach-1979}, ZM model
\cite{Zykov-1986,Loskutov-1990} etc.): there are only three
discrete levels, and relaxation of each CAU is considered as
intralevel transitions (or transitions between virtual sublevels).

\begin{widetext}

The evolution of each individual CAU proceeds by sequential cyclic
transitions $L_K \rightarrow L_{K'}$ (where $K$ and $K' \in \{
{\mathrm{I}}, {\mathrm{III}}, {\mathrm{II}} \}$), induced by the
Kolmogorov evolution operator $\widehat\Omega$
\cite{Kolmogorov-etc}. In the TLCA model, the evolution operator
$\widehat\Omega$ has three orthogonal branches
$\widehat\Omega_{\mathrm{I}}$, $\widehat\Omega_{\mathrm{III}}$ and
$\widehat\Omega_{\mathrm{II}}$, which we call ground, excited and
refractory branches respectively. The choice of the branch at
iteration $n+1$ is dictated only by the global attribute of the CAU
at step $n$, namely:

\begin{equation}
  \varphi_{ij}^{(n+1)} = \widehat\Omega\left(\varphi_{ij}^{(n)}\right) =
  \begin{cases}
    \widehat\Omega_{\mathrm{I}}\left(\varphi_{ij}^{(n)}\right),  & \text{if $\varphi_{ij}^{(n)} = 0$},\\
    \widehat\Omega_{\mathrm{III}}\left(\varphi_{ij}^{(n)}\right),& \text{if $0 < \varphi_{ij}^{(n)} \leq \tau_e$},\\
    \widehat\Omega_{\mathrm{II}}\left(\varphi_{ij}^{(n)}\right), & \text{if $\tau_e < \varphi_{ij}^{(n)} \leq \tau_e + \tau_r$}.
  \end{cases}
  \label{eq:17}
\end{equation}

\subsection{Ground branch of the evolution operator}\label{subse:iii-B}

At step $n+1$, the branch $\widehat\Omega_{\mathrm{I}}$ by
definition fulfills operations only over those CAUs, which have
$L_K = L_{\mathrm{I}}$ at step $n$. These operations are precisely
the same, as in ZM model \cite{Zykov-1986,Loskutov-1990}, namely:

\begin{equation}
  \varphi_{ij}^{(n+1)} = \widehat\Omega_{\mathrm{I}}\left(\varphi_{ij}^{(n)}\right) =
  \begin{cases}
    0, & \text{if $\bigl( \varphi_{ij}^{(n)} = 0 \bigr) \wedge \bigl( u_{ij}^{(n+1)} < h \bigr)$};\\
    1, & \text{if $\bigl( \varphi_{ij}^{(n)} = 0 \bigr) \wedge \bigl( u_{ij}^{(n+1)} \geq h \bigr)$},
  \end{cases}
  \label{eq:18}
\end{equation}

\vspace{12pt}

\begin{equation}
  u_{ij}^{(n+1)} =
  A_{ij}^{(n)} \left( S_{ij}^{(n)} \right) + D_{ij}^{(n)} \left( S_{i+p,j+q}^{(n)} \right) =
  g u_{ij}^{(n)} + \sum_{p,q} C(p,q)J_{i+p,j+q}^{(n)},
  \label{eq:19}
\end{equation}
where $A_{ij}^{(n)}$ is the accumulating term for the
$u$\nobreakdash-\hspace{0pt}agent, which is an analog of chemical
activator \cite{Zykov-1986,Loskutov-1990}; $D_{ij}^{(n)}$ is the
first-channel diffusion term; $h$ is the threshold for the
$u$\nobreakdash-\hspace{0pt}agent ($h
> 0$); $g$ is the accumulation factor for the
$u$\nobreakdash-\hspace{0pt}agent ($g \in [0,1])$; $C(p,q)$ is the
active neighborhood of the CAU at site $(i,j)$; and the
$u$\nobreakdash-\hspace{0pt}agent arrives to CAUs with $L_K =
L_{\mathrm{I}}$ only from CAUs with $L_K = L_{\mathrm{III}}$ in
$C(p,q)$:
\begin{equation}
  J_{i+p,j+q}^{(n)} =
  \begin{cases}
    1, & \text{if $\bigl( 0 < \varphi_{i+p,j+q}^{(n)} \leq \tau_e \bigr)$};\\
    0, & \text{if $\bigl( \varphi_{i+p,j+q}^{(n)} > \tau_e \bigr) \vee \bigl( \varphi_{i+p,j+q}^{(n)} = 0 \bigr)$}.
  \end{cases}
  \label{eq:20}
\end{equation}

The definition of the diffusion term $D_{ij}^{(n)}$ in Eqn.
(\ref{eq:19}) is very flexible. Apart of well-known neighborhoods
of the Moore ($C(p,q) = C_M(p,q)$) and the von Neumann ($C(p,q) =
C_N(p,q)$) types:

\begin{equation}
  C_M(p,q) =
  \begin{cases}
    1, & \text{if $\Bigl[ \bigl( \, |p| \leq 1 \bigr) \wedge \bigl( \, |q| \leq 1 \bigr) \Bigr] \wedge \bigl( \delta_{p0}\delta_{q0} \neq 1 \bigr)$};\\
    0, & \text{otherwise},
  \end{cases}
  \label{eq:21}
\end{equation}
\begin{equation}
  C_N(p,q) =
  \begin{cases}
    1, & \text{if $\Bigl[ \bigl( \, |r| = 1 \bigr) \oplus \bigl( \, |s| = 1 \bigr) \Bigr]$};\\
    0, & \text{otherwise},
  \end{cases}
  \label{eq:22}
\end{equation}
one can easy define another neighborhoods, e.~g. of the box type
$C_B(p,q)$ or of the diamond type $C_D(p,q)$ \cite{Fisch-1991}
(which are the straightforward generalization of the  Moore
$C_M(p,q)$ and the von Neumann $C_N(p,q)$ neighborhoods), etc.

On the other hand, the ZM definition
\cite{Zykov-1986,Loskutov-1990} of the weight factors $
J_{i+p,j+q}^{(n)}$ (\ref{eq:20}) may be extended out of the binary
set $\{0,1\}$ to model various distance-dependent phenomena within,
e.~g., $C_D(p,q), C_B(p,q)$.

In this work, however, we restricted ourselves by the Moore
neighborhood and by weight factors of the form (\ref{eq:20}).
Another types of neighborhoods, weight factors and some other
modifications of the model will be studied in subsequent papers.

\subsection{Excited branch of the evolution operator}\label{subse:iii-C}

At the same step $n+1$, the branch $\widehat\Omega_{\mathrm{III}}$
fulfills operations only over those CAUs, which have $L_K =
L_{\mathrm{III}}$ at step $n$:
\begin{equation}
  \varphi_{ij}^{(n+1)} =
  \widehat\Omega_{\mathrm{III}}\left( \varphi_{ij}^{(n)} \right)=
  \begin{cases}
    \varphi_{ij}^{(n)} + 1, & \text{if $\left[ \bigl( 0 < \varphi_{ij}^{(n)} < \tau_e \bigr) \wedge \bigl( z_{i,j}^{(n+1)} < f \bigr) \right] \vee \bigl( \varphi_{i,j}^{(n)} = \tau_e \bigr)$};\\
    \varphi_{ij}^{(n)} + 2, & \text{if $\bigl( 0 < \varphi_{ij}^{(n)} < \tau_e \bigr) \wedge \bigl( z_{i,j}^{(n+1)} \geq f \bigr)$}.
  \end{cases}
  \label{eq:23}
\end{equation}
\begin{equation}
  z_{ij}^{(n+1)} =
  \overline{D}_{ij}^{\:(n)} \left( S_{i+p,j+q}^{(n)} \right) =
  \sum_{p,q} C(p,q)Q_{i+p,j+q}^{(n)}.
  \label{eq:24}
\end{equation}
where $\overline{D}_{ij}^{\:(n)}$ is the second-channel diffusion
term; $f$ is the threshold for the
$z$\nobreakdash-\hspace{0pt}agent ($f>0$), and we assume that
$z$\nobreakdash-\hspace{0pt}agent arrives to excited CAU ($L_K =
L_{\mathrm{III}}$) only from those CAUs, which have $L_K =
L_{\mathrm{I}}$ in $C(p,q)$:
\begin{equation}
  Q_{i+p,j+q}^{(n)} =
  \begin{cases}
    1, & \text{if $\varphi_{i+p,j+q}^{(n)} = 0$};\\
    0, & \text{if $\varphi_{i+p,j+q}^{(n)} \neq 0$}.
  \end{cases}
  \label{eq:25}
\end{equation}

One can see from (\ref{eq:24}) that the
$z$\nobreakdash-\hspace{0pt}agent does not accumulate during
successive iterations. In other words, the branch
$\widehat\Omega_{\mathrm{III}}$ at step $n+1$ produces "memoryless"
values of partial attributes $z_{ij}^{(n+1)}$ for CAUs having $L_K
= L_{\mathrm{III}}$ at step $n$ (in contrary to the
$u$\nobreakdash-\hspace{0pt}agent for CAUs having $L_K =
L_{\mathrm{I}}$ at step $n$). The $z$\nobreakdash-\hspace{0pt}agent
may accelerate transitions from excited CAUs to refractory ones.
This is important difference between 2C model and the original 1C
model of ZM \cite{Zykov-1986,Loskutov-1990}; and 2C model
qualitatively describes the real active media used in quantum
acoustics (see Appendices \ref{ap:A} and \ref{ap:B}).

\subsection{Refractory branch of the evolution operator}\label{subse:iii-D}

The branch $\widehat\Omega_{\mathrm{II}}$ does not produce/change
any partial attributes at all (because the intermediate level
$L_{\mathrm{II}}$ is in the state of refractority). It fulfils only
such the operations over CAUs, which have $L_K = L_{\mathrm{II}}$
at step $n$:
\begin{equation}
  \varphi_{ij}^{(n+1)} =
  \widehat\Omega_{\mathrm{II}}\left(\varphi_{ij}^{(n)}\right)=
  \begin{cases}
    \varphi_{ij}^{(n)} + 1, & \text{if $\tau_e < \varphi_{ij}^{(n)} < \tau_e + \tau_r$};\\
                         0, & \text{if $\varphi_{ij}^{(n)} = \tau_e + \tau_r$}.
  \end{cases}
  \label{eq:26}
\end{equation}

Generally speaking, there are many examples of active media with
weak refractority (when the unit at the intermediate level
$L_{\mathrm{II}}$ is not absolutely isolated from its
neighborhooding units). But in this work we restrict ourselves by
the case of perfect refractority (\ref{eq:26}), which is valid,
e.~g., for the phaser systems of the ${\mathrm{Ni^{2+}:Al_2O_3}}$
type (as it was pointed out in Chapter \ref{se:i} and is explained
in Appendix \ref{ap:A}).

\subsection{Reduction of the TLCA model to the ZM automaton}\label{subse:iii-E}

The second-channel diffusion gives the contribution to the TLCA
dynamics if $f \leq {\mathcal{N}}$, i.e. if $f \leq 4$ for $C =
C_N$, $f \leq 8$ for $C = C_M$ and so on.

At $\bigl ( (C = C_N) \wedge (f > 4) \bigr )$ our TLCA model is of
ZM-like (i.e. 1C) type, and at $\bigl ( (C = C_M) \wedge (f > 8)
\bigr )$ it becomes equivalent to the original ZM model
\cite{Zykov-1986,Loskutov-1990}, which (with slight rearrangement
of cases) is as follows:
\begin{equation}
  \varphi_{ij}^{(n+1)} =
  \begin{cases}
    \varphi_{ij}^{(n)} + 1, & \text{if $0 < \varphi_{ij}^{(n)} < \tau_e + \tau_r$};\\
    1, & \text{if $\bigl( \varphi_{ij}^{(n)} = 0 \bigr) \wedge \bigl( u_{ij}^{(n+1)} \geq h \bigr)$};\\
    0, & \text{if $\left[ \bigl( \varphi_{ij}^{(n)} = 0 \bigr) \wedge \bigl( u_{ij}^{(n+1)} < h \bigr) \right] \vee \bigl( \varphi_{ij}^{(n)} = \tau_e + \tau_r \bigr)$},
  \end{cases}
  \label{eq:27}
\end{equation}
where $u_{ij}^{(n)}$ is defined as in (\ref{eq:19})--(\ref{eq:21}).

\end{widetext}

\subsection{Comparison with other models}\label{subse:iii-F}

In our computer experiments we used such the conditions: $\bigl(
J_{i+p,j+q}^{(n)} \in \{0;1\} \bigr) \wedge \bigl( C(p,q) \in
\{0;1\} \bigr)$. Under these conditions and in particular (but very
important) case of $g \in \{0;1\}$, the ZM model
\cite{Zykov-1986,Loskutov-1990} becomes of fully integer kind. If
all these conditions are supplemented by $Q_{i+p,j+q}^{(n)} \in
\{0;1\}$, the TLCA model (\ref{eq:11})--(\ref{eq:26}) becomes fully
integer too. Finally, at $C = C_M, g = 0, h = 1, f > 8$ the TLCA
model (\ref{eq:11})--(\ref{eq:26}) is equivalent to the simplest
variant of the ZM model \cite{Zykov-1986,Loskutov-1990}, which
corresponds to a discrete form of the original WR model
\cite{Wiener-1946}.

On the other hand, the TLCA model (\ref{eq:11})--(\ref{eq:26})
(and, usually, the original ZM model
\cite{Zykov-1986,Loskutov-1990}) may be considered as the
generalization of the SSR model. This generalization is different
from those of R.~Fisch~e.~a. \cite{Fisch-1991} in many aspects;
first of them consists in using a concept of multilevel transitions
in \cite{Fisch-1991} instead of a concept of relaxation times, used
in works \cite{Zykov-1986,Loskutov-1990} and in the present work,
formulae (\ref{eq:11}) to (\ref{eq:26}). But in the limiting case
of ``instant'' (i.e. one-step) transitions both TLCA model and the
model of R.~Fisch~e.~a. \cite{Fisch-1991} reduce to the same
original SSR model. For the TLCA model this is the case of $C =
C_N, f > 4, g = 0, h = 1, \tau_e = \tau_r = 1$. Under these
conditions the system of equations (\ref{eq:11})--(\ref{eq:26})
reduces to (\ref{eq:03})--(\ref{eq:08}) if one uses the ansatz
$\bigl( \phi_{ij}^{(n)}  \in \{-1,0,1\} \bigr) \rightarrow \bigl(
\phi_{ij}^{(n)} \in \{2,0,1\} \bigr)$ in the SSR model. Such the
ansatz is correct because the SSR model in the form of
(\ref{eq:03})--(\ref{eq:08}) is almost insensitive to the phase
counter value for $L_{\mathrm{II}}$: the ``$-1$'' in the equations
(\ref{eq:05}) and (\ref{eq:07}) may be changed to any $X \not\in
\{0,+1\}$.

At $(\tau_e > 1) \vee (\tau_r > 1)$ our TLCA model is of MSR type,
and relaxation of each {\textit{individual}} CAU is described in
the same manner as in various Reshodko's MSR models of excitable
media \cite{Bogach-1979}, but diffusion mechanism in TLCA model is
algorithmized using more versatile way.

There is no direct correspondence between TLCA (ZM) model and the
OK2 model introduced in \cite{Malinetskii-1992}, despite of
definite similarities in spatio-temporal dynamics of these CA.
Partial correspondence is between TLCA (ZM) model and one of
variants of Toffoli-Margolus model, namely with its
``monotonic-threshold'' variant (see Chap.~9 in
\cite{Toffoli-1987}). The TLCA model reduces to that variant of
Toffoli-Margolus model if $C = C_M, f
> 8, g = 0, h \in \{2,3,4\}, \tau_e = 1, \tau_r = 2$.

\subsection{Geometry of active media, boundary conditions and transients in CA}\label{subse:iii-G}

Pattern evolution is very sensitive to geometry of active medium
and boundary conditions even in the framework of operation of the
same CA. There are three types of geometry and boundary condition
(GBC) combinations most commonly used in computer experiments with
CA (see Table~I). Let us consider them in details.

$\bullet$ The GBC\nobreakdash-\hspace{0pt}1 type is defined as
follows:
\begin{equation}
  S^{(n)} \bigl( i_{\mathrm{v}},j_{\mathrm{v}} \bigr) =
  S^{(n)} \bigl( i_{\mathrm{e}}^{\Gamma},j_{\mathrm{e}}^{\Gamma} \bigr),
\label{eq:28}
\end{equation}
where
\begin{equation}
  i_{\mathrm{e}}^{\Gamma} =
    \begin{cases}
    M_X             & \text{if $i_{\mathrm{v}} = 0$}; \\
    1,              & \text{if $i_{\mathrm{v}} = M_X + 1$}; \\
    i_{\mathrm{v}}, & \text{otherwise},
  \end{cases}
  \label{eq:29}
\end{equation}
\begin{equation}
  j_{\mathrm{e}}^{\Gamma} =
    \begin{cases}
    M_Y             & \text{if $j_{\mathrm{v}} = 0$}; \\
    1,              & \text{if $j_{\mathrm{v}} = M_Y + 1$}; \\
    j_{\mathrm{v}}, & \text{otherwise}.
  \end{cases}
  \label{eq:30}
\end{equation}
Here $(i_{\mathrm{v}},j_{\mathrm{v}}) \in
{\mathfrak{M}}_{\mathrm{v}}$, and the coordinates
$(i_{\mathrm{e}}^{\Gamma},j_{\mathrm{e}}^{\Gamma})$ belong to the
border set ${\mathfrak{M}}_{\mathrm{e}}^{\Gamma} \subset
{\mathfrak{M}}_{\mathrm{e}}$ of excitable area, i.~e.
$(i_{\mathrm{e}},j_{\mathrm{e}}) =
(i_{\mathrm{e}}^{\Gamma},j_{\mathrm{e}}^{\Gamma}$) if $(i = 1) \vee
(j = 1) \vee (i = M_X) \vee (j = M_Y)$.

\vspace{10pt}

$\bullet$ The GBC\nobreakdash-\hspace{0pt}2 type can be easily
defined by using an extended interval for the phase counter
$\varphi$. Let every cell $(i_{\mathrm{v}},j_{\mathrm{v}})$ in
${\mathfrak{M}}_{\mathrm{v}}$ contains one {\textit{unexcitable}}
CAU. All these unexcitable CAUs have frozen $\varphi^{(n)} \bigl(
i_{\mathrm{v}},j_{\mathrm{v}} \bigr)$ at all $n \in [0,N]$ without
reference to states of
$(i_{\mathrm{e}}^{\Gamma},j_{\mathrm{e}}^{\Gamma})$, and
\begin{equation}
  \varphi^{(n)} \bigl( i_{\mathrm{v}},j_{\mathrm{v}} \bigr) = \tau_e + \tau_r + a,
  \label{eq:31}
\end{equation}
where, $a > 0$, say $a = 1$ without loss of generality. These
unexcitable CAUs in ${\mathfrak{M}}_{\mathrm{v}}$ ``absorb''
excitations at the border of ${\mathfrak{M}}_{\mathrm{e}}$ in
contrary to the case of GBC\nobreakdash-\hspace{0pt}1, where
excitations at the border ${\mathfrak{M}}_{\mathrm{e}}$ are
reinjected in active medium. In many cases this difference leads to
qualitatively different behaviour of the whole CA. Note that by
this way one may also define CA with {\textit{inhomogeneous}}
active medium (where some of CAUs are unexcitable not only in
${\mathfrak{M}}_{\mathrm{v}}$, but in ${\mathfrak{M}}_{\mathrm{e}}$
too), introducing an additional orthogonal branch
$\widehat\Omega_0$ into the evolution operator $\widehat\Omega$.
This branch is activated at $\varphi_{ij}^{(n)} = \tau_e + \tau_r +
1$ only and is simply the identity operator $\widehat\Omega_0 =
\widehat{I}$, i.e.: $\varphi_{ij}^{(n+1)} = \widehat\Omega_0
\varphi_{ij}^{(n)} = \varphi_{ij}^{(n)}$. And even more complex
behaviour of such ``impurity'' CAUs may be defined using analogous
approach (i.~e. by introducing additional orthogonal branches in an
evolution operator): there may be pacemakers \cite{Loskutov-1990}
or another special units embedded in active medium and described by
phase counter in extended areas $(\varphi >
\tau_e + \tau_r) \vee (\varphi < 0)$.%

$\bullet$ The GBC\nobreakdash-\hspace{0pt}3 type is, strictly
speaking, a case of borderless system. But the starting pattern
has, of course, bounded quantity of CAUs with $L_K \neq
L_{\mathrm{I}}$, located in bounded part of active medium. Hence,
during the whole evolution ($1 \leq n \leq N$) the front of growing
excited area will meet the unexcited (but excitable!) CAUs only.
For the case GBC\nobreakdash-\hspace{0pt}3, there is neither
absorption of excitations at boundaries (as for
GBC\nobreakdash-\hspace{0pt}2), nor feedback by reinjection of
excitations in the active system (as for
GBC\nobreakdash-\hspace{0pt}1). From this point of view CAs of
GBC\nobreakdash-\hspace{0pt}3 type are ``simpler'', than CAs of
GBC\nobreakdash-\hspace{0pt}1 and GBC\nobreakdash-\hspace{0pt}2
types. On the other hand, CAs of GBC\nobreakdash-\hspace{0pt}3 type
are potentially infinite discrete systems where true aperiodic
(irregular, chaotic) motions are possible, in contrast to finite
discrete systems, possessing, of course, only periodic trajectories
in certain phase space after ending of transient stage
\cite{Lorenz-etc}.

\begin{table}
\begin{center}
  \caption{Geometry of active media and boundary
  conditions}
  \begin{tabular}{|c|c|c|} \hline
  {\textit{Type}} & {\textit{Geometry}} & {\textit{Boundary conditions}} \\ \hline
  GBC\nobreakdash-\hspace{0pt}1            & Bounded, Toroidal                    & Cyclic    \\ \hline
  GBC\nobreakdash-\hspace{0pt}2            & Bounded, Flat                        & Zero-flow \\ \hline
  GBC\nobreakdash-\hspace{0pt}3            & Unbounded, Flat                      & (free space) \\ \hline
  \end{tabular}
  \label{ta:01}
\end{center}
\end{table}

On the other hand, there is also very important difference between
GBC\nobreakdash-\hspace{0pt}2 and other two types of GBC. The
system of GBC\nobreakdash-\hspace{0pt}2 type is the single from the
three ones under consideration which interacts with the external
world dynamically (besides of relaxation). Sure enough,
GBC\nobreakdash-\hspace{0pt}1 and GBC\nobreakdash-\hspace{0pt}3 are
connected to this outer world {\textit{only}} by relaxation
channels (the $\tau_e$ and $\tau_r$ are the measures of this
connection). In contrary, a CA of the GBC\nobreakdash-\hspace{0pt}2
type interacts with surroundings through the real boundary, which
is in fact absent in toroidal finite-size active medium of
GBC\nobreakdash-\hspace{0pt}1 type and it is absent by definition
for CA of GBC\nobreakdash-\hspace{0pt}3 type. Despite of elementary
mode of such interaction (boundary simply ''absorbs'' the
outside-directed flow of excitations from
$(i_{\mathrm{e}}^{\Gamma},j_{\mathrm{e}}^{\Gamma})$, a CA of the
GBC\nobreakdash-\hspace{0pt}2 type may demonstrate very special
behaviour. The main attention in this work is devoted just to TLCA
of GBC\nobreakdash-\hspace{0pt}2 type as the most realistic model
of phaser active system, where both the mechanisms of interaction
(dynamical and relaxational) of an active medium with the outer
world are essential. Dissipation in a phaser active medium (highly
perfect single crystal at liquid helium temperatures) is caused by
two main mechanisms: (a) dynamical, by coherent microwave phonon
and photon emission directly through crystal boundaries and (b)
relaxational, by thermal phonon emission. There are, of course,
many more or less important differences between the TLCA model and
the real phaser systems
\cite{Phaser-ruby-early,Nickel-early,DNM-Diss-1983,TPL-2001,arx-selforg-2003,TP-2004,arx04a-destab-2004,Peterson-1969,Tucker-etc},
but in any case autonomous phaser has only these two mechanisms of
interaction with the outer world.

\subsection{Initial conditions}\label{subse:iii-H}

Patterns ${\mathrm{P}}^{(n)}$ in CA are usually described in terms
of levels (or ``colors'') of CAUs, and initial conditions
${\mathrm{P}}^{(0)}$ are formulated simply as matrix of CAUs levels
$L_K$. But states of CAUs in such the automata as ZM or TLCA are
fully defined not only by levels $L_K$ itself. There are global
attributes which must be predefined before TLCA evolution is
started. Some of parial attributes must be predefined too. These
points are essential for reproducing of the results of computer
experiments.

In our work such the initial conditions for the global attributes
$\varphi_{ij}$ are used:
\begin{equation}
  \begin{cases}
    \varphi_{ij}^{(0)} = 0,          & \text{if $L_K^{(0)} = L_{\mathrm{I}}$};\\
    \varphi_{ij}^{(0)} = 1,          & \text{if $L_K^{(0)} = L_{\mathrm{III}}$};\\
    \varphi_{ij}^{(0)} = \tau_e + 1, & \text{if $L_K^{(0)} = L_{\mathrm{II}}$},
  \end{cases}
  \label{eq:32}
\end{equation}
Initial conditions for $u_{ij}$ must be defined for ground-state
CAU's ($L_K^{(0)} = L_{\mathrm{I}}$) only. In this work, we suppose
\begin{equation}
  \bigl( u_{ij}^{(0)} = 0 \bigr) \quad \text{IFF} \quad
  \bigl( \varphi_{ij}^{(0)} = 0 \bigr),
  \label{eq:33}
\end{equation}
where ${\mathrm{IFF}}$ means ``if and only if''. Initial conditions
for $z_{ij}$ are undefined for all $L_K^{(0)}$ because
$z$\nobreakdash-\hspace{0pt}agent is not defined at $n=0$, see Eqn.
(\ref{eq:12}). As the result, starting pattern ${\mathrm{P}}^{(0)}$
may be defined (both for the ZM and TLCA models) as the matrix $ \|
\varphi_{ij}^{(0)} \| $ (where $ \varphi_{ij}^{(0)} \in \{0, 1,
\tau_e + 1 \} $) with additional condition given by Eq.
(\ref{eq:33}).

\section{RESULTS OF COMPUTER EXPERIMENTS}\label{se:iv}

Cellular automata are inherently based on irreducible algorithms.
Generally speaking, there are no predictive procedures for such the
systems. So, the best way to investigate cellular automaton (TLCA
in particular) is to run it, because, as S.~Wolfram pointed out,
"their own evolution is effectively the most efficient procedure
for determining their future'' (see \cite{Wolfram-1985}, page 737).

Here we present results of our computer experiments both with
original 1C automaton of ZM and with 2C automaton described in
Section \ref{se:iii}. The main tool in these experiments was the
program ``Three-Level Cellular Automaton''
(TLCA~\textcircled{c}~2004~S.~D.~Makovetskiy
\cite{SDM-TLCA-package-2004}), which is included now in the
software package ``Three-Level Laser Model'' \cite{SDM-Forum-2005}.
Original ZM-algorithm \cite{Zykov-1986,Loskutov-1990}, was
implemented in a previous software package ``Generalized
Wiener-Rosenblueth Model''
(GWR~\textcircled{c}~2001-2003~S.~D.~Makovetskiy
\cite{SDM-GWR-package-2003}; see also \cite{SDM-VIRT-2002}).

Computer experiments were fulfilled for wide ranges of control
parameters (CP): $(1 \leq \tau_e \leq 50) \wedge (1 \leq \tau_r
\leq 50)$ (for all the cases); $0 < g < 1$ (for non-integer
versions of TLCA and GWR); $g \in \{ 0;1 \}$ (for integer versions
of TLCA and GWR); $2 \leq h \leq 100$ (for all the cases); $1 \leq
f \leq 9$ (the case $f = 9$ corresponds to original ZM model ---
see Subsection \ref{subse:iii-E}).

A discrete system with finite set of levels may have only two types
of dynamically stable states at bounded lattice. The first of them
is stationary state, and the second is periodic one. They may be
called attractors by analogy with lumped dynamical systems (see
e.~g. \cite{Liu-2002,Wuensche-1998,Wuensche-2003}). For our
cellular automaton, the first of such the attractors is
spatially-uniform and time-independent state with
$\varphi_{ij}^{(n)} = 0$, where $(i \in [1,M_X]) \wedge (j \in
[1,M_Y])$, $n \geq n_C$. In other words, the only stationary state
of TLCA is the state of full collapse of excitations at some step
$n_C$ (by definition of excitable system). The second type of TLCA
attractors includes many various periodically repeated states (RSWs
is a typical but not the single case). In this case
$\varphi_{ij}^{(n)} = \varphi_{ij}^{(n+T)}$ where $(i \in [1,M_X])
\wedge (j \in [1,M_Y])$, $n \geq n_P$; $n_P$ is the first step of
motion at a periodic attractor; $T$ is the integer-number period of
this motion, $T > 1$.

If starting patterns ${\mathrm{P}}_{ID}^{(0)}$ (where $ID$ is the
pattern identifier) are generated with random spatial distribution
by levels, then the time intervals $n_C$ or $n_P$ may be considered
as times of full ordering in the system. Such {\textit{irregular}}
transients are of special interest from the point of view of
nonlinear dynamics of distributed systems
\cite{Awazu-2003,Crutchfield-1988,Hramov-2004,Morita-2003,Morita-2004}
because they may be bottlenecked by very slow, intermittent
morphogenesis of spatial-temporal structures. In the next
Subsection we study this collective relaxation both for cases of
collapse and periodic final states of TLCA evolution.

\subsection{Bottlenecked collective relaxation in TLCA}\label{subse:iv-A-Bottlenecked}

Evolution of TLCA usually has several stages with very different
characteristic times and qualitatively different spatio-temporal
dynamics. An example of relatively simple evolution is shown at
Figure~1 for starting pattern ${\mathrm{P}}^{(0)} = {\mathrm{P}}^{(0)}_A$.%
\footnote{Typical set of starting patterns with $M_X = M_Y = 100$,
used in our numerical experiments, see Figure~A1.}
Here $f = 8$, but at $f = 9$ this starting pattern reaches
precisely the same attractor. This is because our TLCA model at $f
= 8$ is very close to ZM model. But even at $f = 8$ such coinciding
of attractors is relatively rare (less than 10 percent for the CP
set pointed at caption to Figure~1 and starting patterns of the
type shown at Figure~A1). In other words TLCA model even at $f = 8$
{\textit{is not equal}} to ZM model. Much more typical case is
gradual divergence of spatio-temporal structures for$f = 8$ and $f
= 9$ during evolution (see Figure~2). And for $f < 8$ we have not
observed any attractors coinciding with ones for $f = 9$.

Evolution of TLCA with $f < 8$ is more complicated, and transient
time may rich giant values (millions of steps for the same $M_X =
M_Y = 100$ as for patterns at Figure~1 and Figure~2) due to
bottlenecked RSWs morphogenesis. Under certain conditions, the
morphogenesis of RSWs proceeds by multiple irregular changings of
effective topolgical charge%
\footnote{Effective topolgical charge is defined in the near
vicinity of the vortex core.}
$Q_T$, including reversing of ${\mathrm{sgn}}(Q_T)$. For example,
at $f = 4$ the system with starting pattern ${\mathrm{P}}^{(0)} =
{\mathrm{P}}^{(0)}_C$ (see the ${\mathrm{P}}^{(0)}_C$ at Figure~A1)
reaches its periodic attractor only at $n \approx 4 \cdot 10^6$.
Seemingly perfect RSW with $Q_T = +2$ is forming at $n \approx 3
\cdot 10^5$, but it is obviously not an attractor. At $n \approx
4.5 \cdot 10^5$ the core of RSW becomes complicated and gradually
evolves to state with $Q_T = +3$ (Figure~3, step $n = 500525$).
During sequence of several metamorphoses, the system returns to
$Q_T = +2$ (Figure~3, step $n = 1.35 \cdot 10^6 )$. And at $n =
1.63 \cdot 10^6$ one can see RSW with opposite sign ($Q_T = -2$).
The further evolution continues this unpredictable scenario, ending
at attractor with $Q_T = -3$ (steps $n = 3.97 \cdot 10^6$ to $n = 4
\cdot 10^6$ at Figure~3). Besides of quantity and sign of $Q_T$,
the frequency of spatial waves irrregularly changes too ---
compare, e.~g., patterns at $n = 2 \cdot 10^5$ and $n = 1.09 \cdot
10^6$ (or at $n = 1.94 \cdot 10^6$ and $n = 2.59 \cdot 10^6$ etc.).

It is interesting to compare such scenario of evolution of TLCA
with a scenario of crystal growth due to spiral (screw) defects
increasing. The results are just opposite: crystal growth is
usually accelerated by many orders due to screw defects, but TLCA
relaxation (transient to an attractor) is highly bottlenecked by
RSWs.

There is a huge amount of periodic attractors for such TLCA, but
the steady-state attractor is the single one. It is fully collapsed
state of an excitable system, as it was pointed out earlier. On the
other hand, there is very many starting patterns that evolve to
this collapsed state (our TLCA is an irreversible system). An
example of very slow transition to collapse is shown at Figure~4.

Bottlenecked collective relaxation shown at Figure~3 and Figure~4
is, of course, aperiodic. But it has some quasiperiodic features
because observed at Figure~3 and Figure~4 spirals (or spiral
``domains'') possess more or less regular structure. Much more
chaotic spatio-temporal behaviour one can observe varying CP of the
TLCA. Such a complex, slow and unpredictable evolution of a simple
TLCA system (with $f = 9$) is shown at Figure~5. This is a typical
example of transient spatio-temporal chaos in collective relaxation
of excitations, which we observed in TLCA.

Such highly bottlenecked and irregular transient dynamics
(metaphorically, ``turbulence'') was observed numerically by
J.~P.~Crutchfield and K.~Kaneko \cite{Crutchfield-1988} in an 1D
model which is a kind of coupled map lattice (CML) model
\cite{Kaneko-1986}. In contrary to CA and other discrete mappings,
CML is not fully discrete system because of continuous spectrum of
states of its elementary units. In other words, CML has not any
well resolved levels. But both TLCA and CML demonstrate lethargic
transients which are not consequences of usual critical slowing
down (i.~e. singularity of transient time at several combination of
CP \cite{Gorban-etc}). Critical slowing down takes place near
critical points only, and fine tuning is needed to reach the (very
narrow) band of CP where this type of slowing-down is observable.

In phaser {\textit{amplifier}} the observed phenomenon of critical
slowing down \cite{RiE-2004} may be described by simple lumped
model of inversion states, see Appendix \ref{ap:C}. But in
autonomous phaser {\textit{generator}} we experimentally observed
\cite{Tashkent-1991-etc} phenomena of spin-phonon interaction with
very long, lethargic aperiodic transients (phonon ``turbulence'')
which have qualitatively another nature than usual critical slowing
down.

There are some qualitatively different mechanisms of slowing down,
for which such the tuning is not necessary. A well known mechanism
of this type is the effect of self-organized criticality
\cite{Bak-etc}, which takes place when a dissipative system holds
themself in critical state and no external tuning of CP is needed
(models of sandpiles, earthquakes, forest fire etc.).

Some interesting scenarios of bottlenecked relaxation were proposed
recently by K.~Kaneko and his co-authors
\cite{Awazu-2003,Morita-2003,Morita-2004}. In particular,
self-organized bottleneck was revealed by H.~Morita and K.~Kaneko
\cite{Morita-2003} in transient processes for simple excited
Hamiltonian system. Here ``the critical state is
{\textit{spontaneously}} formed without continuous driving, once a
{\textit{part}} of a system is highly excited'' \cite{Morita-2003}.
And in work of A.~Awazu and K.~Kaneko \cite{Awazu-2003} it was
found that forming and evolution of transient dissipative
structures lead to non-critical slowing down in closed chemical
system (a modified Brusselator model).

Super-slow evolution of seemingly ``frozen'' spiral 2D structures
was revealed and investigated by another group (C.~Brito,
I.~S.~Aranson, and H.~Chat\'e) \cite{Brito-2002}) in 2D system
modeled by complex Ginzburg-Landau equation. During last years some
new publications on slow evolving continuous systems has been
appeared, particularly in the ``Condensed Matter'' and ``Nonlinear
Sciences'' arXives.

Collective relaxation phenomena observed by us in TLCA are of
similar kind despite of {\textit{fully dicrete}} nature of our
model. Bottlenecked transition to final state of TLCA caused by
RSWs may qualitatively explain why relaxation times of
level-populations in dissipative three-level (or multilevel)
paramagnetic systems differ by several orders from those predicted
by one-particle spin-lattice relaxation theory
\cite{Altshuler-1972,Faughnan-1961} (this old problem has not
satisfactory solution up to now). On the other hand, freezing,
quasi-freezing and slow evolution are now a subject of
investigation of three-level systems in many alternative directions
of modern nonlinear science, see e.~g. work of F.~Vazquez,
P.~L.~Krapivsky, and S.~Redner \cite{Vazquez-2002} etc.

Hypersensitivity of the TLCA to changing of initial conditions
(which will be discussed in the next Subsections of this work),
instability of vortices and extremely slow, lethargic moving to a
regular attractor are typical features of transient deterministic
chaotic behaviour of TLCA. Birth, evolution, interaction and decay
of RSWs in TLCA (including original ZM model), as it is obvious
from Figures 1 to 5, is much more complex and unpredictable than it
was expected at early days of investigation of such axiomatic CA
models of excitable systems. In particular, collisions of RSWs with
absorbing borders (in the GBC-2 case) violate conservation of $Q_T$
(see Figure~3 and Figure~5). In the next Subsection we describe
some our numerical results concerning the problem of competition of
left- and right-handed spatio-temporal structures, which is of
interest not only from physical point of view
\cite{Brandenburg-2004}.

\subsection{Competition of left- and right-handed RSWs}\label{subse:iv-LeftRight}

A bounded solitary domain of excited CAUs having appropriate
relaxation times and placed far from grid boundaries (or at
unbounded grid) may evolve to RSWs if and only if there is an
adjoined (but not a surrounding) domain of refractory CAUs
\cite{Balakhovskii-1965,Loskutov-1990,Fisch-1991,Selfridge-1948,Toffoli-1987}.
In this case RSWs apear by pair with opposite ${\mathrm{sgn}}
(Q_T)$ and integral $Q_T^{(\text{integr})} \equiv \sum_m Q_T^{(m)}$
is obviously conserved (by infinity in time if grid is unbounded).
If such solitary excited-refractory area is placed in a starting
pattern near the grid boundary and the GBC-2 conditions take place,
the single RSW appears and evolves and $Q_T^{(\text{integr})}$ is,
of course, not conserved in this case. These simplest and well
known examples illustrate possibilities of coexistence and
competition of one or two RSWs in excitable media, and possible
scenarios of the RSW(s) evolution may be easily forecasted.

An evolution of complex patterns with multiple, irregularly
appearing, chaotically-like drifting and colliding RSWs (or,
possibly, another spatio-temporal structures posessing handedness)
is in essence umpredictable without direct computing of the whole
transient stage. In particular, some intuitive predictions about of
$Q_T^{(\text{integr})}$ at attractor may be fully errorneous, as it
will be shown in this Subsection.

In the previous Subsection, it was already pointed out that, for
some CP sets, an effective topological charge $Q_T$ (more
precisely, $Q_T^{(\text{integr})}$) irregularly changes, including
reversing of its sign during a system evolution. Extremely high
complexity of almost all pre-attractor evolution (see e.~g.
Figure~5) does not permit to prognose even the near future of such
TLCA. But sometimes it seems that long-term prediction is yet
possible for another CP sets. In particular, Figure~6 demonstrate
such ``possibility'' at least at pre-finish stage of transient
process in TLCA with $f = 3$ (other CP are the same as for
Figures~1 to 4) and with $M_X = M_Y = 300$. Indeed, beginning from
$n > 10^5$ the sign of the integral effective topological charge
becomes equal to $\mathrm{sgn} \left( Q_T^{(\text{integr})} \right)
= - 1$ (Figure~6, $n= 1.6 \cdot 10^5$) and stays unchanged up to
winning of the single RSW in the spiral competitions%
. In other word, a ``perfect'' left-handed word is
formed here.

Nevertheless, this pretty picture of ``predictable''
self-organization is fully destroyed at $n \approx 3.96 \cdot 10^5$
due to collision of the winner's core with the grid boundary and
subsequent full collapse of excitations at $n_C = 396982$. This
last stage is not shown at Figure~6 because it is of the same type
as collapse at Figures~4 and 5 -- it is simply a gradual absorbing
of non-spiral waves of excitation by boundaries of our GBC-2
system.

But much more unexpected, fully counter-intuitive TLCA dynamics may
be illustrated by two snapshots (see Figures~7 and 8) of evolution
of a system with the same CP set as for Figure~6, but with $M_X =
M_Y = 900$. At Figure~7, one can see almost purely right-handed
pattern at step $n = 1.48 \cdot 10^6$ of TLCA evolution. The only
deviation from ideal right-handedness at Figures~7 is the
rudimentary counter-clockwise ``tail'' of the clockwise spiral at
top-center part of Figure~7 (note that this highly asymmetric
spiral-antispiral pair is inside of the closed wavefront).

One may expect that the winner RSW in this system will be of the
same handedness (by analogy with the Figure~6, where left-handed
multi-spiral pattern was evolved to left-handed solitary RSW). But
the subsequent evolution of the system is very surprising, see
Figure~8. The winner RSW is shown here at step $n = 2 \cdot 10^6$,
it rotates counter-clockwise ($Q_T = -1$), i.~e. the system
{\textit{becomes left-handed despite of their long right-handed
previous life}}. During further evolution ($n > 2 \cdot 10^6$, not
shown here), the winner's core is moving irregularly across the
grid up to collision with the grid boundary, and full collapse of
excitations takes place at $n_C = 2260964$ with final stage by
scenario shown at Figures~4 and 5.

We will emphasise that the single \emph{left-handed} RSW (Figure~8)
is the result of evolution of the \emph{right-handed} pattern of
RSWs (Figure~7). A series of intermediate snapshots (not shown here
due to their big size) at the time interval from $n = 1.48 \cdot
10^6$ to $n = 2 \cdot 10^6$ demonstrate a scenario of such
reversing of the system handedness. An the cause of this reversing
is not rudimentary counter-clockwise ``tail'' of the clockwise
spiral at top-center part of Figure~7, because it quickly dies
during ``spiral war''. Moreover, at $n = 1.88 \cdot 10^6$ the
single fully developed right-handed RSW remains at the whole grid,
in the right-bottom corner of the grid. The second competing
right-handed RSW is in a dangerous proximity from left border of
the grid at the same $n = 1.88 \cdot 10^6$. And a small dislocation
in spiral wavefront of the first RSW appears as the nucleus of
future left-handed RSW.

At $n = 1.89 \cdot 10^6$, the two competing right-handed RSWs are
still alive (both of them are near corners of grid), but the
mentioned dislocation has already transormed to fully developed
left-handed RSW. During the subsequent evolution, both right-handed
RSWs are died, the remnant of wavefronts of right-handed structure
is crowded out by giant lacuna, and the left-handed RSW becomes the
winner in the world with almost perfect right-handed history.
Finally, this solitary left-handed RSW must die too at $n_C \approx
2.26 \cdot 10^6$, as it was pointed already.

This intriguing story confirms high level of unpredictability of
the system under consideration, and this subject will be discussed
in details below (see Section \ref{se:v}). And now we describe
another unusual form of motions in our TLCA observed at different
values $f = 5$ of the threshold for the

\subsection{Chimera states - coexistence of regular and chaotic domains in
TLCA}\label{subse:iv-Chimera}

At the same CP $\tau_e, \tau_r, h, g$ as in Figures~1 to 4 and
Figures~6 to 8, but with $f = 5$ we revealed unusual, long-living
stages of CA evolution, during which coexistence of periodic and
aperiodic spatio-temporal structures takes place in TLCA. A useful
tool for investigation of such the phenomenon is generalized
Poincar\'{e} cross-section of pattern sequence.

Ascending and descending generalized Poincar\'{e} cross-sections,
used in this work, are defined as follows. Let ${\mathrm{B}}(n)$ is
an 1-bit (e.~g. black = TRUE, white = FALSE) pattern and $\Delta n$
is a selection interval ($\Delta n > 1$). For each binary cell
${\mathrm{B}}_{ij} \in {\mathrm{B}}$, we fulfill the sequence of
$k$ logical additions in the ascending time (starting from a step
$n_1 \leq N - k \Delta n$):
\begin{equation}
 \begin{split}
  {{\mathrm{B}}^{(+)}_{ij}}(n_1, \Delta n, k) =
  & [\:[\:[\:{\mathrm{B}}_{ij}(n_1) \vee \\
  & \vee {\mathrm{B}}_{ij}(n_1 + \Delta n)\:] \vee \\
  & \vee {\mathrm{B}}_{ij}(n_1 + 2\Delta n\:)] \vee ... \\
  & \vee {\mathrm{B}}_{ij}(n_1 + k\Delta n)\:],
 \label{eq:34}
 \end{split}
 \end{equation}
and a sequence of $m$ logical additions in the descending time
(starting from a step $n_2 > m \Delta n$):
\begin{equation}
 \begin{split}
  {{\mathrm{B}}^{(-)}_{ij}}(n_2, \Delta n, m) =
  & [\:[\:[\:{\mathrm{B}}_{ij}(n_2) \vee \\
  & \vee {\mathrm{B}}_{ij}(n_2 - \Delta n)\:] \vee \\
  & \vee {\mathrm{B}}_{ij}(n_2 - 2\Delta n)\:] \vee ... \\
  & \vee {\mathrm{B}}_{ij}(n_2 - m\Delta n)\:].
 \label{eq:35}
 \end{split}
 \end{equation}

Ascending and descending generalized Poincar\'{e} cross-sections
(${\mathcal{P}}^{(+)}$ and ${\mathcal{P}}^{(-)}$) are the patterns,
which consist of ${\mathrm{B}}^{(+)}_{ij}(n_1, \Delta n, k)$ and
${\mathrm{B}}^{(-)}_{ij}(n_2, \Delta n, m)$ respectively:
\begin{equation}
 {\mathcal{P}}^{(+)}(n_1, \Delta n, k)
 = \| {\mathrm{B}}^{(+)}_{ij}(n_1, \Delta n, k) \|
 \label{eq:36}
\end{equation}
\begin{equation}
 {\mathcal{P}}^{(-)}(n_2, \Delta n, m)
 = \| {\mathrm{B}}^{(-)}_{ij}(n_2, \Delta n, m) \|
 \label{eq:37}
\end{equation}

Figure~9 demonstrates phenomenon of coexistence of long-living
periodic and aperiodic spatio-temporal structures in TLCA by using
of ascending and descending generalized Poincar\'{e} cross-sections
(left and right columns of Figure~9 respectively; the central
column at Figure~9 is usual sequence of TLCA patterns showed for
clarity). We use term ``transient chimera states'' for such
structures by analogy to term introduced by D.~M.~Abrams and
S.~H.~Strogatz \cite{Abrams-2004} for analogous non-transient
phenomena revealed earlier by Y.~Kuramoto and D.~Battogtokh
\cite{Kuramoto-2002} (see below in this Subsection).

The series of ascending generalized Poincar\'{e} cross-sections
${\mathcal{P}}^{(+)} (n_1, \Delta n, k_p)$ shows that spatial
domain of aperiodic motions is bounded at the whole interval of
selection. In other words, periodic domain is dynamically stable
(beginning at list from $1.5 \cdot 10^5$) for more than half of the
grid area. And the series of descending generalized Poincar\'{e}
cross-sections ${\mathcal{P}}^{(-)}(n_2, \Delta n, m_q)$
demonstrates growing of the regular domain, i.~e. how the system
reaches the attractor keeping previously occupied territory of
local periodicity. This evolving coexistence of two qualitatively
different (periodically and aperiodically oscillating) domains in
high-dimensional discrete mapping we call transient chimera states.

For $\tau_e = \tau_r = h = 50$, $g =1$ and $f \neq 5$ reaching of
an attractor proceeds typically by metamorphoses of aperiodic waves
to periodic ones over the whole grid, without coexistence of
periodic and aperiodic domains. But for some alternative values of
$\tau_e, \tau_r, h, g$ and for $f \neq 5$ we observed chimera
states again. Evolution of some of these states are different
comparatively to ones shown at Figure~9. For example, small but
strictly regular spatial islands in chaotic sea may appear; they
live by long time and then dissappear during that or those stages
of TLCA evolution; and attractor may be reached after disappearing
of all of these islands. In other words, emergence of regular
domain(s) of CAUs (shown at Figure~9) is not the single scenario of
an attractor reaching when spatially-periodic and spatially-chaotic
states are coexisting.

Similar phenomena of coexistence of incongruous or even
antagonistic spatio-temporal structures were discovered during last
years in computer experiments with 1D arrays of identical coupled
oscillators \cite{Abrams-2004,Kuramoto-2002} and 2D grids emulating
blackouts in electric power systems \cite{Carreras-etc}. The most
essential difference between systems, which was used in
\cite{Abrams-2004,Carreras-etc,Kuramoto-2002} and in the present
work, consists in the nature of their elementary units. In works
\cite{Abrams-2004,Kuramoto-2002} and in overview
\cite{Carreras-etc} one deals with units, which by definition
oscillate even without any coupling with their neighborhoods. In
the TLCA model (as well as in all models of excitable media) each
isolated elementary unit has the single stable state, namely steady
ground state.

In other words, the TLCA or another system of this kind may be
treated as oscillator if (and only if) there is more or less strong
interaction between elementary units. And namely the whole TLCA is
an oscillator in this case. Electric power generators or another
monochromatic oscillators, of course, may work autonomously ---
with their own frequencies and phases. But a grid with
noninteracting excitable CAUs at $n > \tau_e +\tau_r$ is simply a
collection of anabiotic cells. So the concepts of phase
synchronization/desynchronization are unapplicable to individual
CAUs by definition. But synchronization may take place between
domains of CAUs (and this is not a trivial ``synchronization'' of
e.g. pacemaker-like domains in excitable media with stable domain
walls).

To conclude this Subsection, we mention that coexistence of
{\textit{static}} regular and disordered structures is well known
in CGL (complex ''still life'' configurations of CAUs). E.~g., some
unbounded (GBC-3 conditions) two-level ($L_G \in \{ L_0, L_1 \}$)
CGL-like automata demonstrate unlimited growing of domain(s) with
frozen CAU states (Turing structures), were static
spatially-chaotic ``cloud'' is irregularly grided by static
spatially-periodic ``rays''. The simplest CA of this type is the
INKSPOT automaton \cite{Toffoli-1987}, which may be defined
similarly to CGL automaton (see Eqns. (\ref{eq:01}) and
(\ref{eq:02})), using binary phase counter $\Phi_{ij}^{(n)} \in
\{0,1\}$ and the following elementary upgrading rules:
\begin{equation}
  \Phi_{ij}^{(n+1)} =
  \begin{cases}
    \Phi_{ij}^{(n)}, & \text{if $U_{ij}^{(n+1)} \neq 3$};\\
    1,               & \text{if $U_{ij}^{(n+1)} = 3$},
  \end{cases}
  \label{eq:38}
\end{equation}
where $U_{ij}^{(n+1)}$ is defined by Eqn. (\ref{eq:02}) and the
Moore neighborhood is used.

Only $L_0 \rightarrow L_1$ transitions are permitted in the INKSPOT
automaton (\ref{eq:38}), but this restriction is not critical for
the phenomenon of coexisting of regular and chaotic static domains.
A similar behavior demonstrate, e.~g., the RtC (``Rays through
Clouds'') two-level automaton \cite{RiE-1999-etc} defined by rules
\begin{equation}
  \Phi_{ij}^{(n+1)} =
  \begin{cases}
    \Phi_{ij}^{(n)}, & \text{if $\left( U_{ij}^{(n+1)} = 2 \right) \vee \left( U_{ij}^{(n+1)} \in [4;8] \right) $};\\
    1,               & \text{if $U_{ij}^{(n+1)} = 3$};\\
    0,               & \text{otherwise},
  \end{cases}
  \label{eq:39}
\end{equation}
or, in more convenient form:
\begin{equation}
  \Phi_{ij}^{(n+1)} =
  \begin{cases}
    \Phi_{ij}^{(n)}, & \text{if $U_{ij}^{(n+1)} \notin \{ 0; 1; 3 \}$};\\
    1,               & \text{if $U_{ij}^{(n+1)} = 3$};\\
    0,               & \text{if $U_{ij}^{(n+1)} \in \{ 0; 1 \}$},
  \end{cases}
  \label{eq:40}
\end{equation}
It is clear from the equation (\ref{eq:40}) that both transitions
$L_0 \rightleftarrows L_1$ are permitted for the RtC automaton, but
emergent Turing structures of chimera type is still the main form
of evolution of this CA (except of special starting patterns giving
pure rays) \cite{RiE-1999-etc}.

\subsection{Spatial symmetry breaking and restoring}\label{subse:iv-Symmetry}

Computer experiments with spatially random input leave out an
another important case of initialization of a system by symmetric
or almost symmetric starting pattern. In this Subsection we
investigate TLCA evolution with such (almost) symmetric initial
conditions.

For a perfectly symmetric input, the TLCA evolution will proceed by
conserved-symmetry scenario. It directly follows from the TLCA
definition as homogeneous and isotropic (in the von Neumann sense
\cite{vonNeumann-1966}) Kolmogorov machine with deterministic
rules. But how will the TLCA evolve if a perfect symmetry of input
pattern is more or less distorted? Will symmetry be broken out or
it may be restored? Or both these scenarios are possible due to
multistability of TLCA? One cannot find answers to these questions
{\textit{a priori}}, using only general guidelines as Curie's
principle \cite{Brading-2003}.

To find such the answers empirically, let us study TLCA evolution
with strictly determined asymmetry of staring patterns (Figure~10).

Left-column starting pattern ${\mathrm{P}}_{ML}^{(0)}$ at Figure~10
is not fully symmetric (ten black cells at the top of the excited
left part of ${\mathrm{P}}_{ML}^{(0)}$ are changed to ten
{\textit{white}} cells at the top of the excited right part
${\mathrm{P}}_{ML}^{(0)}$). But the $C_v$ symmetry is fully
restored already at $n = 48$ (not shown at Figure~10), and at $n =
10129$  the system is already at periodic attractor (``Smile of
Cheshire cat''
\footnote{This is, of course, not a ``Cheshire cat'' of
C.~R.~Tompkins (a two-level block predecessor, which is known in
CGL).}
 --- see Figure~10, left column, row $d$). This attractor has
period $T = 3592$ and consists of two mirror-symmetric,
counter-rotating RSW with $| Q_T | = 3$.

The right-column starting pattern ${\mathrm{P}}_{MR}^{(0)}$ at
Figure~10 differs from the left-column starting pattern
${\mathrm{P}}_{ML}^{(0)}$ only by the single CAU with coordinates
$i = 66$; $j = 41$ (cells are numbered beginning from the left
bottom corner of the grid). But evolution of ${\mathrm{P}}_{MR}$ is
qualitatively different from the ${\mathrm{P}}_{ML}$ case. At $n =
500$ (right column, row $b$ at Figure~10) ${\mathrm{P}}_{MR}$ is
fairly asymmetric. At $n = 1256$ one can see that
${\mathrm{P}}_{MR}$ takes form of a RSW with $Q_T = +4$ (right
column, row $c$ at Figure~10). But this RSW is unstable --- the
attractor for ${\mathrm{P}}_{MR}$ is RSW with $Q_T = +3$ and period
$T = 2308$ (right column, row $d$ at Figure~10).

This example shows that hypersensitivity of TLCA to changing of
initial conditions may lead not only to full ``breaking'' of the
slightly imperfect symmetry, but to restoring of such imperfect
symmetry to the perfect one. Besides of $C_v$ or $C_h$ cases, a
similar competition of symmetry ``breaking'' and restoring is
possible for the cases of $C_2$, $C_4$ (both under GBC-1 or GBC-2
conditions). If starting patterns consists of multiple symmetric
parts, then similar phenomena will appear at least under GBC-1
conditions (for perfect mosaic pattern and the GBC-1 conditions, a
regular vortex grid is obviosly possible).

Besides of these broken/restored symmetry phenomena, there are
another group of effects, which are connected with competition of
left/right spinning of RSWs. We already described such phenomena in
Subsection \ref{subse:iv-LeftRight} for cases of random starting
patterns. One can mistakenly conclude that unpredictability of
direction of RSW spinning is the consequence of randomness of
inputs in general. But in the next Subsection we will show that the
main source of unpredictability of handedness in TLCA is
hypersensitivity to initial conditions, when minimal possible
change of starting pattern may reverse direction of RSW spinning
(both at attractor and at transient stages of evolution).

\subsection{Hypersensitivity to initial condition and
unpredictability of left- or right-handed forms of
vorticity}\label{subse:iv-Hypersensitivity-left-right}

Changing of a single CAU may lead to strongly expressed mutation of
spatio-temporal structures in TLCA, as it was demonstrated in the
previous Subsection. Such hypersensitivity is an inherent property
of very wide class of nonlinear multicomponent systems with growing
spatial disturbances and multiple attractors. Some more or less
close analogs one can find in such well-known area of technology as
electric power grids, which already has been mentioned within the
framework of chimera states consideration in Subsection
\ref{subse:iv-Chimera} (on cascading effects in power grids and
other real-world and articicial multicomponent systems see also
\cite{Kinney-2004} and references therein).

Now we will investigate mutations which produce reversing of
handedness of attractors reached during TLCA evolution in
conditions of hypersebsitivity to initial conditions. Let the
configurations of {\textit{ground-state}} CAUs in starting patterns
at Figure~11 (${\mathrm{P}}_{SL}^{(0)}$ at top of left column and
${\mathrm{P}}_{SR}^{(0)}$ at top of right column) are perfectly
symmetric (in distinction to starting patterns
${\mathrm{P}}_{ML}^{(0)}$ and ${\mathrm{P}}_{MR}^{(0)}$ shown at
Figure~10). So asymmetry in both ${\mathrm{P}}_{SL}^{(0)}$ and
${\mathrm{P}}_{SR}^{(0)}$ (Figure~11) is determined only by
slightly differentiated populations of two non-ground levels.

Namely, left-column starting pattern ${\mathrm{P}}_{SL}^{(0)}$ at
Figure~11 is not fully mirror-symmetric, because ten black cells at
the top of the excited left part of ${\mathrm{P}}_{SL}^{(0)}$ are
changed to ten {\textit{gray}} cells at the top of the excited
right part ${\mathrm{P}}_{SL}^{(0)}$. At $n = 2010$ pattern
${\mathrm{P}}_{SL}$ takes form of a RSW with $Q_T = +4$ (left
column, row $c$ at Figure~11). But this RSW is unstable --- the
attractor for ${\mathrm{P}}_{MR}$ is RSW with $Q_T = +3$ (left
column, row $d$ at Figure~11).

The right-column starting pattern ${\mathrm{P}}_{SR}^{(0)}$ at
Figure~11 differs from the left-column starting pattern
${\mathrm{P}}_{SL}^{(0)}$ only by the single CAU with coordinates
$i = 66$; $j = 41$ (similarly to that for Figure~10, but black cell
at site $(66; 41)$ is changed by gray one). Evolution of
${\mathrm{P}}_{SR}$ is qualitatively different from the
${\mathrm{P}}_{SL}$ case and leads to an attractor having another $
| Q_T | $ and opposite ${\mathrm{sgn}} (Q_T) $, namely to RSW with
$Q_T = -2$ (right column, row $d$ at Figure~11).

This last result illustrates that hypersensitivity of TLCA to
changing of initial conditions is a cause of unpredictability of
the direction of a RSW spinning even for the robust vortices.
Minimal possible disturbance of the initial conditions (by
changhing of the state of a single CAU in a starting pattern)
trigger the sign of $Q_T$, not to speak of the $Q_T$ modulo. In
other words, a result of self-organization of a dynamically stable
world in TLCA cannot be forecasted even for almost identical
initial conditions. Left-handed or right-handed life in the
framework of the model under consideration is a result of pure
accident despite of extrmely simple, absolutely deterministic
underlying algorithm.

In conclusion of this computer-experiment Section it will be marked
out the following. Rich phenomenology of TLCA evolution,
illustrated in this Section by several important examples
concerning RSWs
\footnote{Besides of RSWs, there are several qualitatively another
types of robust spatio-temporal structures in TLCA, and a lot of
fragile ones. They will be discussed in separate publication.}
, demonstrates that this kind of deterministic mapping is one of
the simplest, fully discrete, bounded (except the case GBC-3)
system possessing complexity of very high degree. Brief list of
observed complex phenomena includes:

$\bullet$ Transient spatio-temporal chaos (including long-term
irregular evolution of complicated RSW structures);

$\bullet$ Competition of left-handed and right-handed RSWs with
chaotic alterations of $Q_T^{(\text{integr})}$ and unexpected
handedness of the winner RSW;

$\bullet$ Incongruous or fully antagonistic spatio-temporal
structures (chimerae);

$\bullet$ Amazing symmetry-connected properties of TLCA caused by
hypersensitivity to initial condition, including full restoring of
(initially broken) symmetry of evolving TLCA pattern. Reversing of
${\text{sgn}} \left ( Q_T^{(\text{integr})} \right )$ with changing
of $| Q_T^{(\text{integr})} |$ (mutations) caused by modification
of a single CAU state only at the beginning of evolution.

In the Section \ref{se:v} we discuss relationship between
simplicity of the TLCA model and complexity of its behaviour as
well as some other issues concerning self-organization in CA.

\section{DISCUSSION}\label{se:v}

A model of an active medium based on cyclic transitions in a
three-level system was introduced for the first time by
astrophysist A.~D.~Thackerey in 1930's \cite{Thackerey-etc}. In
essence, this was a model of resonatorless laser system, but
proposed about 20 years before the beginning of the laser era (and
about 10 years before work of N.~Wiener and A.~Rosenblueth
\cite{Wiener-1946} on excitable systems and work of J.~von~Neumann
\cite{vonNeumann-1966} on cellular automata).

Similarity of class-B lasers and excitable (chemical) systems was
revealed and investigated by C.~O.~Weiss, K.~Staliunas and their
co-workers \cite{Staliunas-1995,Weiss-1993,Weiss-1999,Weiss-2004}
without using of CA or any other discrete mapping of this kind. Are
algorithmically simple (but having phase space of huge dimension)
CA models of class-B lasers and/or laser-like systems appropriate
for at least qualitative study of these active devices? Are there
advantages of CA in modeling of laser (phaser) dynamics at all?

There are many publications on CA and other algorithmically simple
models of complex systems, which may be joined under the common
title: ``Complex worlds from simple rules?'' (this is the real
title of the {\textit{shortest}}, to our knowledge, paper on this
subject --- see note of U.~Yurtsever \cite{Yurtsever-2002}).
Criticism of such publications is directed against simplicity of
CAU operation, but it usually does not take into account emergent
{\textit{co-operation}} of large or very large quantity of CAUs in
an intrinsically unstable system with random (or randomly
perturbed) initial conditions.

It is obvious that discrete, spatially bounded deterministic
system, e.~g. CA (ZM or TLCA in particular) may have only two types
of attractors, both of which are regular (as it was already
stressed in Section \ref{se:iv}). In countrary, continuous
deterministic system may have a great variety of irregular
attractors \cite{Schuster-1984} or even much more complicated
(turbulent) spatio-temporal structures%
\footnote{Some intermediate cases are represented by discrete,
{\textit{spatially unbounded}} systems (e.g. CA of GBC-3 type, see
Section \ref{se:iii}), where true irregular spatio-temporal
structures may appear.}
. Algorithmic complexity of models simulating continuous systems is
usually higher (by many reasons, including purely technical).
Introducing of stochastic terms in a (classic) model brings new
sources of complexity. And quantum-mechanical models in most cases
has the highest level of complexity. This hierarchy of complexity
is built by the principle of {\textit{algorithmic}} complexity. But
algorithmic or Kolmogorov complexity is not the unique tool and it
possibly is not an adequate tool in the area under consideration
\cite{Shalizi-2003}. So (in countrary to the U.~Yurtsever
\cite{Yurtsever-2002} point of view)  ``cellular automata are one
of the more popular and distinctive classes of models of complex
systems'' (see \cite{Shalizi-2003}, page 20).

Results of the present work demonstrate complex, unpredictable
behaviour of algorithmically simple system containing mesoscopic
quantity of discrete elementary units (usually $10^4 - 10^6$). We
would like to stress that we deal with high-dimensional system,
which cannot be reduced to any averaged, low-dimensional one.
Distinctions between ``standard'' system having dimension of phase
space $d_{\mathrm{phase}} = 3$ and a system with
$d_{\mathrm{phase}} = 10^4 - 10^6$ is, of course, very significant
circumstance \cite{Albers-2004} in modeling of active systems
containing discrete interacting particles. Phaser system is one of
such systems. Real phaser contains macroscopic quantity of
particles ($d_{\mathrm{phase}} > 10^{16} - 10^{17}$)
\cite{DNM-Diss-1983,Tashkent-1991-etc,TPL-2001,arx-selforg-2003,TP-2004,arx04a-destab-2004},
but we hope that mesoscopic system is more appropriate for modeling
of spatio-temporal chaotic (``turbulent'') motions
\cite{Tashkent-1991-etc}, self-organization and slow self-detunings
\cite{TPL-2001,arx-selforg-2003,TP-2004,arx04a-destab-2004},
inversion collapse \cite{Nickel-early,DNM-Diss-1983} and other
nonlinear phenomena in this multi-particle dissipative system%
\footnote{Phenomena observed in driven phasers
\cite{Nickel-early,DNM-Diss-1983,TPL-2001,arx-selforg-2003,TP-2004,arx04a-destab-2004}
need modification of our TLCA model to a non-autonomous one.}.

In particular, transient spatio-temporal chaos observed in TLCA may
be collated with complex, ``turbulent'' behaviour of autonomous
phaser generator \cite{Tashkent-1991-etc}. On the other hand,
experimentally revealed coexistence of regular and irregular
motions in ruby phaser \cite{RiE-1999-etc} may be interpreted as
chimera states \cite{Abrams-2004,Kuramoto-2002} observed in TLCA as
transient phenomenon (see Figure~9).

Generally, TLCA model describes oscillatory collective states in
system of non-oscillatory CAUs, including local
self-synchronization of oscillations caused by diffusion of
excitations, and global synchronization (or, alternatively, full
collapse of excitations) as final state of the system evolution. In
the first case a robust RSW or more complicated but strictly
periodic in time structure is formed being a cyclic attractor of
TLCA at several initial conditions --- see e.~g. Figure~3. This is
an analog of limit cycle, i. e. regular attractor known in lumped
dynamical systems. In the second case, when all non-ground CAU
break out at finish, the full freezing of TLCA takes place --- see
e.~g. Figure~4. This is an analog of point-like attractor in lumped
dynamical systems. Both attractors may coexist for the same set of
CP, as it is for the cases at Figure~3 and Figure~4. Such
coexistence of final states is not the full analog of generalized
multistability, which is well known for lumped dynamical systems,
but branching of self-organized states in TLCA is evident.

It is interesting to discuss shortly the question of degree of
self-organization in the systems of CA type \cite{Shalizi-2004}. It
is obvious intuitively that final robust RSW in a TLCA (Figure~3)
is a result of high-degree self-organization in TLCA. But
{\textit{it is not obvious intuitively}} that absolutely frozen
final state (Figure~4) of TLCA is self-organized. At the same time,
both routes to final states at these figures are very similar. This
or that attractor is reached due to predefined (but deeply hidden)
route dependent on combination of TLCA rules, control parameters
and initial conditions. Slight changing of initial conditions
frequently leads to another attractor, but the evolution itself may
by qualitatively the same. If evolution is highly bottlenecked then
attractor may not be reached at all during the whole time of
numerical or real experiment.

From this point of view a final state itself may be of minor
interest, more important is a transient dynamics. The work of
J.~P.~Crutchfield and K.~Kaneko \cite{Crutchfield-1988}, which was
already cited in Section \ref{se:iv}, was titled: ``Are Attractors
Relevant to Turbulence?''. We may slightly reformulate this
question in the context of our study: ``Are Attractors Relevant to
Transient Spatio-Temporal Chaos?''. The answer is ``Yes'' if an
attractor may be reached for a reasonable time $t_{\text{attr}}$
(in fully discrete and bounded system $t_{\text{attr}}$ is always
limited by quantity of all possible states of the system). But the
answer is ``No'' if $t_{\text{attr}}$ exceeds any possible duration
of an experiment. In this case a system with transient
spatio-temporal chaos cannot be distinguished from true chaotic
system without additional testing.

As a matter of fact, there are some intermediate classes of
phenomena ``at the edge between order and chaos'' which may appear
in bounded discrete deterministic system with large phase space.
And self-organization scenario which includes super-slow,
bottlenecked, chaotic-like stages is a signature of dominance of
such an intermediate class of system dynamics in numerical
experiments. It is important to define not only qualitative
criteria of self-organization, but quantitative ones too. Really,
having limited time and computer capacity, one cannot reach final
self-organized state for a system with huge dimension of phase
space. Irreducibility of CA or another discrete mappings does not
permit direct forecasting of the system future without direct
computation. So the computed part of transient process is the
single source of available information of our fully deterministic
but partially determined system.

Criteria of self-organization introduced during last decades are of
great interest. The brightest of them, absolutely counter-intuitive
example is suggestion of Yu.~L.~Klimontovich
\cite{Klimontovich-1996}, which may be formulated shortly as
follows: Turbulence is more organized state than laminar motion.
This suggestion is based on using of normalized entropy of
nonequilibrium state. Paradoxical approach of Klimontovich is
actually deep and have definite perspective for development in
nonequilibrium thermodynamics. But it is inadequate for very wide
class of (self-)organized complex system including high-dimensional
system of CA type.

Nonequilibrium thermodynamics is very capricious thing, it may give
both excellent and unappropriate results for the same physical
object under different conditions. Our previous experiments on
phaser systems confirmed the need of careful analysis of
applicability of thermodynamical approach to concrete cases. Very
good qualitative and even quantitative interpretation of
experiments on phaser amplification is contrasted to the case of
phaser generation, where spin-temperature model sometimes gives
very bad description of phenomena observed (and such models are
failed to produce any heuristics of phaser generation). In an
amplifier, there are no self-organized states at all, and saturated
populations of levels admit averaged description (by spin
temperature in rotating frame) \cite{Phaser-ruby-early}
--- in contrary to some regimes of phaser generator \cite{Tashkent-1991-etc}.
Using of exact high-dimensional CA models is, to our mind, more
appropriate for investigation of self-organized multiparticle
systems.

Recently some interesting results were reported on quantitative
investigation of self-organization in cyclic CA
\cite{Shalizi-2004}, which are a kind of multilevel CA
\cite{Fisch-1991} with SSR (see Section \ref{se:ii}). Cyclic CA are
more expedient for chemical systems, which haven't real discrete
levels. Phasers (as far other active systems based on inversion of
energy-level populations), in contrary to chemical excitable
systems, have well resolved, discrete levels (usually three or
four), so TLCA is an adequate model from this point of view.
Moreover, our TLCA is of MSR type, so it is a flexible tool taking
into account relaxation times for nonground levels. Application of
the approach, proposed in \cite{Shalizi-2004}, to the TLCA may give
a useful tool for direct quantitative study of self-organization in
this cellular automaton and, possibly, for understanding of
nonlinear phenomena observed in real physical experiments with
Gigahertz phaser generator
\cite{DNM-Diss-1983,Tashkent-1991-etc,TP-2004,arx04a-destab-2004}
and similar active systems with negligible level of quantum noise.

And a bit about applicability of the TLCA to modeling of
optical-range phonon lasers must be explained. Several successful
attempts of laser-like generation of phonons at low-frequency side
of Terahertz range were undertaken at the end of 1970-th
\cite{Meltzer-etc}, and now the renewed interest to this frequency
range is coming into sight \cite{Terahertz-new}. There are two main
differences between 10-Gigahertz (usual microwaves) and, say,
100-Terahertz (or 1000-Terahertz) ranges: (i) about 12 (up to 15)
orders difference in level of quantum noises, already disscussed in
Section \ref{se:i}, and (ii) sufficient distinction in distribution
of level-population, which will be considered here for the case of
three-level system.

In an excitable medium of a Gigahertz quantum device, all three
working energy levels (ground, excited and refractory levels in
context of the present work) are populated even without inversion
and even at liquid helium temperatures. Short pulse inversion
almost instantly changes relative distribution of active centers
with different states, but after the inversion pulse ending, this
or that part of active centers remains in the refractory state.
Examples of such distributions are shown at Figure~A1. In this case
the TLCA algorithm is directly applicable for modeling of
spatio-temporal dynamics in active (excitable) medium.

On the other hand, at optical-range frequencies only the ground
level $L_{\mathrm{I}}$ is populated in equilibrium system. Pulse
inversion primary populates the highest level $L_{\mathrm{III}}$,
but there still are no active centers at intermediate (refractory)
level $L_{\mathrm{II}}$ during at least first stage of inversion.
To model dynamics of such the medium correctly, the TLCA must be
complemented by appropriate branch(es) of evolution operator
describing pointed critical stage of population distribution. This
is a future direction of modeling of laser-like active systems by
CA.

\section{CONCLUSIONS}\label{se:vi}

In this work, we fulfill computer modeling of spatio-temporal
dynamics in large (up to $10^6$ particles) dicrete systems of
three-level CAUs interacting by short-range diffusion through one
or two channels. Long-time evolution ($10^5 - 10^7$ iterations) of
TLCA demonstrates some unusual phenomena including highly
bottlenecked collective relaxation of excitations by multiple RSWs
with variable $Q_T$; competition of left- and right-handed RSWs
with alteration of integral $Q_T$; long lived chimera states, i.~e.
coexistence of regular and chaotic domains in evolving patterns of
CAUs; branching of TLCA states with different symmetry which, in
particular, leads to full restoring of symmetry of imperfect
starting pattern. Most of observed processes of TLCA evolution may
be attributed to transient high-dimensional spatio-temporal chaos
in a deterministic excitable system of three-level active units.
Super-slow collective motions, spatio-temporal transient chaos and
coexistence of regular and chaotic structures in TLCA qualitatively
describe part of real experimental data on phaser (microwave phonon
laser) dynamics.

\appendix
\section{ENERGY LEVELS AND MICROWAVE-FREQUENCY TRANSITIONS
IN ${\mathbf{Ni^{2+}:Al_2O_3}}$ SPIN SYSTEM}\label{ap:A}

Free $3d^8$-ion $\mathrm{Ni^{2+}}$ has ground term
${}^3{\mathrm{F}}$ (see, e.~g.
\cite{Abragam-1970,Altshuler-1972,Petrosyan-1986}). This term
consists of $(2{\mathcal{L}} + 1) \times (2{\mathcal{S}} + 1) = 21$
levels, which are unsplitted for a $\mathrm{Ni^{2+}}$ ion in free
space (Figure~A2). Here ${\mathcal{L}}$ and ${\mathcal{S}}$ are the
orbital and the spin quantum numbers respectively: ${\mathcal{L}} =
3$, ${\mathcal{S}} = 1$.

In corundum (${\mathrm{Al_2O_3}}$) crystal matrix doped by nickel,
each ${\mathrm{Ni^{2+}}}$ ion is under action of static
six-coordinated trigonal electric field (called crystal field
\cite{Abragam-1970,Altshuler-1972}) with the highest-symmetry axis
$\bm{\mathcal{O}}_3$. This third-order axis coincides with the
optical axis of corundum crystal matrix. In the Appendices
\ref{ap:A}, \ref{ap:B} we will use the Cartesian coordinate system
with the applicata $\bm{\mathcal{O}}_z$ directed along the axis
$\bm{\mathcal{O}}_3$.

The electric field of the trigonal symmetry splits the ground term
${}^3{\mathrm{F}}$ into orbital singlet ${}^3{\mathrm{A}}_2$ and
two orbital triplets ${}^3{\mathrm{T}}_1$ and ${}^3{\mathrm{T}}_2$
(Figure~A2). The lowest ${}^3{\mathrm{A}}_2$ orbital singlet in
non-relativistic approximation has three unsplitted spin levels
(the orbital singlet ${}^3{\mathrm{A}}_2$ is simultaneously the
spin triplet), because crystal (electric) field in this
approximation does not interact with spin (magnetic) moment.

But this degeneracy is partially takes off by the relativistic
spin-orbit interaction, described by operator $\lambda
\widehat{\bm{\mathcal{L}}} \widehat{\bm{\mathcal{S}}}$
\cite{Abragam-1970,Altshuler-1972}, where $\lambda$ is the constant
of spin-orbital interaction; $\widehat{\bm{\mathcal{L}}}$ and
$\widehat{\bm{\mathcal{S}}}$ are the operators of the orbital and
the spin momenta respectively. As the result, spin triplet
${}^3{\mathrm{A}}_2$ becomes splitted into low-lying spin doublet
and the excited spin singlet (Figure~A2). The spin singlet for the
${\mathrm{Ni^{2+}:Al_2O_3}}$ spin-system lies at ${\mathcal{D}}_0 /
\hbar \approx 39.8$~GHz over the doublet
\cite{Abragam-1970,Altshuler-1972}, where ${\mathcal{D}}_0$ is so
called zero field (more rigorously, zero-magnetic-field) splitting
of quantum energy levels.

This splitting in ${\mathrm{Ni^{2+}:Al_2O_3}}$ is about fourfold
greater than ${\mathcal{D}}_0$ in ruby
(${\mathrm{Cr^{3+}:Al_2O_3}}$) due to sufficient difference in $|
\lambda |$ for ${\mathrm{Ni^{2+}}}$ and ${\mathrm{Cr^{3+}}}$ ions:
$| \lambda ({\mathrm{Ni^{2+}}}) | = 335$~cm${}^{-1}$
{\textit{versus}} $| \lambda ({\mathrm{Cr^{3+}}}) | =
87$~cm${}^{-1}$ \cite{Altshuler-1972}. Accordingly, spin-phonon
interaction in ${\mathrm{Ni^{2+}:Al_2O_3}}$ is much stronger (and
longitudinal paramagnetic relaxation is much faster) than in ruby
--- compare experimental data on tensor of spin-phonon interaction
for ${\mathrm{Ni^{2+}:Al_2O_3}}$ \cite{Nickel-early} and
${\mathrm{Cr^{3+}:Al_2O_3}}$ \cite{Spin-phonon-ruby}). We want to
underline close interconnection between static spin-lattice
interaction (which defines zero-magnetic-field structure of the
low-lying energy levels), dynamical spin-phonon processes (which
respond for the resonant interaction of paramagnetic ions with
microwave ultrasound) and longitudinal relaxation in such the
diluted paramagnetics. The features of longitudinal relaxation of
Nickel ions in ${\mathrm{Al_2O_3}}$ will be consider in the
Appendix \ref{ap:B}. Now we will continue description of the energy
levels and microwave-frequency transitions in the
${\mathrm{Ni^{2+}:Al_2O_3}}$ paramagnetic system.

In nonzero static magnetic field ${\mathbf{H}}$, which is applied
along $\bm{\mathcal{O}}_3$, the doublet splits into pair of Zeeman
levels with $M = \pm 1$, and the three-level spin system
${\mathrm{Ni^{2+}:Al_2O_3}}$ becomes fully splitted (Figure~A2).
The static magnetic field does not affect the level with $M = 0$.
Here $\{M\} \equiv \{-1, 0, +1\}$ is the set of the eigenvalues of
the scalar operator $\widehat{\mathcal{S}}_z$ (this operator is the
projection of the vectorial spin operator
$\widehat{\bm{\mathcal{S}}}$ onto applicata):
\begin{equation}
  \widehat{\mathcal{S}}_z |\Psi_M\rangle = M |\Psi_M\rangle,
  \label{eq:ap-A01}
\end{equation}
where $|\Psi_M\rangle$ are the wave functions (eigenfunctions of
$\widehat{\mathcal{S}}_z$).

The energy levels $E_{(M)}$ of the ${\mathrm{Ni^{2+}:Al_2O_3}}$
spin system for the case under consideration (${\mathbf{H}}
\parallel \bm{\mathcal{O}}_3$) are as follows
\cite{Abragam-1970,Altshuler-1972}:
\begin{gather}
  E_{(0)} = (+2/3) {\mathcal{D}}_0; \label{eq:ap-A02}\\
  E_{(+1)} = (-1/3) {\mathcal{D}}_0 + \hbar {\mathrm{g}}_{\parallel} \beta_B H; \label{eq:ap-A03}\\
  E_{(-1)} = (-1/3) {\mathcal{D}}_0 - \hbar {\mathrm{g}}_{\parallel} \beta_B H, \label{eq:ap-A04}
\end{gather}
where ${\mathrm{g}}_{\parallel}$ is the appropriate component of
the effective g\nobreakdash-\hspace{0pt}factor; $\beta_B$ is the
Bohr magneton; $H = | {\mathbf{H}} |$.

From the practical point of view (e.~g. for using the
${\mathrm{Ni^{2+}:Al_2O_3}}$ as phaser active system
\cite{Nickel-early,DNM-Diss-1983}) the most interest is the case of
small magnetic fields: $H < {\mathcal{D}}_0 / \hbar
\gamma_{\parallel}$. In this case the following correspondence
between ${\mathrm{Ni^{2+}:Al_2O_3}}$ levels $E_{(M)}$ and TLCA
levels $E_K$ (Section \ref{se:iii}) is suggested:
\begin{gather}
  E_{(0)}  \Leftrightarrow L_{\mathrm{III}}; \label{eq:ap-A05}\\
  E_{(+1)} \Leftrightarrow L_{\mathrm{II}}; \label{eq:ap-A06}\\
  E_{(-1)} \Leftrightarrow L_{\mathrm{I}}. \label{eq:ap-A07}
\end{gather}
Possibility of such modeling of the ${\mathrm{Ni^{2+}:Al_2O_3}}$
active spin system by the 2C cellular automaton (Section
\ref{se:iii}) is sustained by the obvious selection rules for
transitions between $E_{(M)}$:
\begin{equation}
  | \Delta M | =
  \begin{cases}
    1, & \text{for $\bigl( E_{(-1)} \rightarrow E_{(0)} \bigr) \wedge \bigl( E_{(0)} \rightarrow E_{(+1)} \bigr)$};\\
    2, & \text{for $\bigl( E_{(+1)} \rightarrow E_{(-1)} \bigr)$}.
  \end{cases}
  \label{eq:ap-A08}
\end{equation}
The main mechanism of diffusion of spin excitations in paramagnetic
systems of the ${\mathrm{Ni^{2+}:Al_2O_3}}$ type is the magnetic
dipole interaction between active units at resonance frequencies
\cite{Abragam-1982,Atsarkin-1978,DNM-Diss-1983}. These interactions
are permitted for $| \Delta M | = 1$, and they are forbidden for $|
\Delta M | = 2$. So, there are only two channels for diffusion of
spin excitations in physical system ${\mathrm{Ni^{2+}:Al_2O_3}}$ at
${\mathbf{H}} \parallel \bm{\mathcal{O}}_3$ --- this is the case of
the diffusion of excitations in the TLCA model (Section
\ref{se:iii}). As the result we have such the correspondence
between $E_{(M)} \rightarrow E_{(M')}$ and $L_K \rightarrow L_{K'}$
\begin{gather}
  \bigl( E_{(0)} \rightarrow E_{(+1)} \bigr)  \Leftrightarrow \bigl( L_{\mathrm{III}} \rightarrow L_{\mathrm{II}} \bigr); \label{eq:ap-A09}\\
  \bigl( E_{(+1)} \rightarrow E_{(-1)} \bigr) \Leftrightarrow \bigl( L_{\mathrm{II}} \rightarrow L_{\mathrm{I}} \bigr);\label{eq:ap-A10}\\
  \bigl( E_{(-1)} \rightarrow E_{(0)} \bigr)  \Leftrightarrow \bigl( L_{\mathrm{I}} \rightarrow L_{\mathrm{III}} \bigr).\label{eq:ap-A11}
\end{gather}
For ${\mathbf{H}} \nparallel \bm{\mathcal{O}}_3$ this close
correspondence is, generally speaking, destroyed. The larger is
angle $\alpha$ between ${\mathbf{H}}$ and $\bm{\mathcal{O}}_3$, the
more mixing of wave functions $|\Psi_M\rangle$ for each level takes
place. But the case of ${\mathbf{H}} \parallel \bm{\mathcal{O}}_3$
(in practice $| \alpha | \lessapprox 5^{\circ}$) is the single
configuration where the phaser generation in the
${\mathrm{Ni^{2+}:Al_2O_3}}$ was experimentally realized
\cite{DNM-Diss-1983}. An attempt of P.~D.~Peterson and
E.~H.~Jacobsen \cite{Peterson-1969} to excite phaser generation in
${\mathrm{Ni^{2+}:Al_2O_3}}$ at $\alpha = 76^{\circ}$ was
unsuccessful despite of large amplification of injected microwave
ultrasound (some possible reasons of this ``silence'' of an active
system ${\mathrm{Ni^{2+}:Al_2O_3}}$ in experiments of
P.~D.~Peterson and E.~H.~Jacobsen \cite{Peterson-1969} were
discussed in \cite{DNM-Diss-1983}). In this work we assume
${\mathbf{H}} \parallel \bm{\mathcal{O}}_3$, and the
${\mathrm{Ni^{2+}:Al_2O_3}}$ system is modeled by TLCA on the basis
of (\ref{eq:ap-A09})--(\ref{eq:ap-A11})

\section{TIMES OF LONGITUDINAL PARAMAGNETIC RELAXATION
IN ${\mathbf{Ni^{2+}:Al_2O_3}}$ SPIN SYSTEM}\label{ap:B}

Most of our early microwave experiments on spin-phonon interaction
and phaser generation in active paramagnetic medium
${\mathrm{Ni^{2+}:Al_2O_3}}$ were fulfilled \cite{DNM-Diss-1983} at
\begin{equation}
  H \ll {\mathcal{D}}_0 / \hbar {\mathrm{g}}_{\parallel} \beta_B,
  \label{eq:ap-B01}
\end{equation}
i.e. under conditions:
\begin{equation}
  F_{(+1,-1)} \ll \bigl( F_{(0,-1)}, F_{(0,+1)}, {\mathcal{D}}_0
  / 2 \pi \hbar \bigr),
  \label{eq:ap-B02}
\end{equation}
where $F_{(M,M')}$ are the frequencies of quantum transitions in
our three-level spin-system:
\begin{equation}
  F_{(M,M')} \equiv \bigl( E_{(M)} - E_{(M')} \bigr) / 2 \pi \hbar
  \label{eq:ap-B03}
\end{equation}
Typical values of CP in experiments
\cite{Nickel-early,DNM-Diss-1983} were as follows: $H \approx
0.5$~kOe; $F_{(+1,-1)} = 3.0$~GHz, $F_{(0,-1)} \approx 41.3$~GHz,
$F_{(0,+1)} \approx 38.3$~GHz. At first blush, such the system may
be described by TLCA with $\tau_e \ll \tau_r$, because time of
direct one-phonon longitudinal relaxation $T_1^{\text{(direct)}}$
(in the {\textit{two-level approximation}} for each the transition)
depends on $F_{(M,M')}$ by such the way
\cite{Abragam-1970,Altshuler-1972}:
\begin{equation}
  T_1^{\text{(direct)}} \propto F_{(M,M')}^{-3}
  \tanh{\bigl( \varkappa F_{(M,M')} / \theta \bigr)}.
  \label{eq:ap-B04}
\end{equation}
Here $\theta$ is temperature of the crystal, $\varkappa$ is
normalizing factor. We use the standard denotation $T_1$ for the
longitudinal relaxation time in physical systems instead of $\tau$,
which is used for relaxation time in the TLCA model system. In this
case $\tau_r$ in the TLCA model (Section \ref{se:iii}) is an analog
of $T_1(E_{(+1)})$, and $\tau_r$ is an analog of $T_1(E_{(0)})$
(this analogy is, of course, only qualitative, but it is clear from
the physical point of view and expedient for numerical modeling of
active systems).

For the case $F_{(M,M')} = F_{(+1,-1)} = 3$~GHz the inequality
$\bigl( \varkappa F_{(M,M')} / \theta \bigr) \ll 1$ takes place at
liquid helium temperatures ($\theta \leq 4.2$~K). So the time of
longitudinal relaxation $T_1^{\text{(direct)}}$ for the level
$E_{(+1)}$ (which is the analog of the refractory level
$L_{\mathrm{II}}$ in TLCA --- see Appendix \ref{ap:A}) at these
temperatures must have such the form:
\begin{equation}
  T_1^{\text{(direct)}} \bigl( E_{(+1)} \bigr) \propto F_{(+1,-1)}^{-2} \theta^{-1}.
  \label{eq:ap-B05}
\end{equation}
This two-level approach (which is often used in modeling of passive
nonlinear systems) leads to wrong conclusion that relaxation of the
level $E_{(0)}$ must be much faster than relaxation of the level
$E_{(+1)}$, and that the correspondent TLCA with $\tau_e \ll
\tau_r$ is a good model for the ${\mathrm{Ni^{2+}:Al_2O_3}}$
physical system
\footnote{Relaxation times $\tau_e$ and $\tau_r$ are, of course,
not the precise analogs of the correspondent physical relaxation
times $T_1 \bigl( E_{(0)} \bigr)$ and $T_1 \bigl( E_{(+1)} \bigr)$
(the last are not additive etc.). Nevertheless, very close
correspondence is obviously present here, and not only qualitative
but quantitative interrelations between physical relaxation times
may be used for defining of $\tau_e$ and $\tau_r$ in a TLCA model
of physical (particularly, phaser) media.}
.

 In a {\textit{three-level}} (or, generally, multilevel) spin
system, especially with large difference in splittings, there
usually are several different mechanisms of longitudinal relaxation
(direct, Orbach, Raman etc. \cite{Abragam-1970,Altshuler-1972}):
\begin{equation}
\begin{split}
  1 / T_1 &= \bigl( 1 / T_1^{\text{(direct)}} \bigr) +\\
  &+ \bigl( 1 / T_1^{\text{(Orbach)}} \bigr) +
  \bigl( 1 / T_1^{\text{(Raman)}} \bigr) + \cdots.
  \label{eq:ap-B06}
\end{split}
\end{equation}
Measurements in ${\mathrm{Ni^{2+}:Al_2O_3}}$ at low temperatures
\cite{Nickel-early,DNM-Diss-1983} showed that the dominant
mechanism of longitudinal relaxation for the spin doublet $M = \pm
1$ at $F_{(+1,-1)} \ll {\mathcal{D}}_0$ is the Orbach two-phonon
process \cite{Abragam-1970,Altshuler-1972}:
\begin{equation}
  T_1 \bigl( E_{(+1)} \bigr) \approx T_1^{\text{(Orbach)}} \bigl( E_{(+1)} \bigr),
  \label{eq:ap-B07}
\end{equation}
with:
\begin{equation}
\begin{split}
  &T_1^{\text{(Orbach)}} \bigl( E_{(+1)} \bigr) \approx
  k_0 {\mathcal{D}}_0^{-3} \exp \bigl( \varkappa {\mathcal{D}}_0
  / \theta \bigr) \approx\\
  &\approx k_0 F_{(0,+1)}^{-3}
  \exp \bigl( \varkappa F_{(0,+1)} / \theta \bigr) \approx
  T_1 \bigl( E_{(0)} \bigr),
  \label{eq:ap-B08}
\end{split}
\end{equation}
where $k_0$ is normalizing factor. As the result, mentioned early
inequality $\tau_e \ll \tau_r$ is generally incorrect for modeling
of dynamics of the ${\mathrm{Ni^{2+}:Al_2O_3}}$ active system (at
least for ${\mathbf{H}} \parallel \bm{\mathcal{O}}_3$, $\theta =
1.8$--$4.2$~K, where the experiments on phaser generation
\cite{Nickel-early,DNM-Diss-1983} were fulfilled). In the case
under consideration, an approximate equality $\tau_e \approx
\tau_r$ takes place instead of the above strong inequality, as it
follows from Eqns. (\ref{eq:ap-B07}) and (\ref{eq:ap-B08}).

Additional changing of physical relaxation times at spin
transitions of ${\mathrm{Ni^{2+}:Al_2O_3}}$ is caused by
interaction (cross-relaxation and some other mechanisms) of
${\mathrm{Ni^{2+}}}$ ions with the Jahn-Teller $3d^7$ ions
${\mathrm{Ni^{3+}}}$ \cite{DNM-Diss-1983} in nickel-doped corundum.
In strong six-coordinated crystal field the single unpaired $3d$
electron of ${\mathrm{Ni^{3+}}}$ ion is at
${\mathrm{e}}$\nobreakdash-\hspace{0pt}orbital, i.~e. the ground
state of ${\mathrm{Ni^{3+}}}$ in corundum is ${}^2{\mathrm{E}}$
(orbital doublet and spin doublet simultaneously) \cite{Shen-etc}.
Non-zero ground state orbital momentum of the ${\mathrm{Ni^{3+}}}$
ion in the ${\mathrm{Ni^{3+}:Al_2O_3}}$ system (in contrast to zero
ground-state orbital momentum for ${\mathrm{Ni^{2+}:Al_2O_3}}$) is
the cause of strong electron-phonon interaction and fast
longitudinal relaxation of the ${\mathrm{Ni^{3+}}}$
\cite{Shen-etc}.

The last type of ions always is present in nickel doped corundum,
because ion ${\mathrm{Ni^{3+}}}$ has the same charge as
${\mathrm{Al^{3+}}}$, and after usual crystal growing the main part
of nickel ions is in trivalent state. Reducing of
${\mathrm{Ni^{3+}}}$ to ${\mathrm{Ni^{2+}}}$ needs special
technological operations, and such a reducing can not be full in
principle.

Interaction of ${\mathrm{Ni^{2+}}}$ with ${\mathrm{Ni^{3+}}}$ in
corundum strongly depends on magnetic resonance frequencies
$F_{(M,M')}(\bm{H})$, this interaction is very sensitive to
concentrations of both ions in corundum, etc. As the result, the
ratios of relaxation times in ${\mathrm{Ni^{2+}:Al_2O_3}}$
spin-system varies in very wide range, depending not only on
experimental conditions, but on a crystal growing technology too.
Due to these circumstances, the adequate TLCA model of the
${\mathrm{Ni^{2+}:Al_2O_3}}$ active system must have possibilities
for working with arbitrary ratios $\tau_e / \tau_r$.

\section{COLLAPSE OF INVERSION STATES AND CRITICAL SLOWING-DOWN\protect\\
IN A THREE-LEVEL PHASER AMPLIFIER\protect\\ WITH BISTABLE
PUMPING}\label{ap:C}

Quantum amplifiers with bistable pumping were primarily
investigated for the simplest case of weak signal
\cite{ZhTF-1991,ZhTF-1999}, when amplified field does not affect
the active medium. The spin system of {\textit{nonlinear}}
three-level phaser amplifier (of a ${\mathrm{Ni^{2+}:Al_2O_3}}$
type \cite{Nickel-early,DNM-Diss-1983,RiE-2003,Cen2ch5-2003}) is
saturated by two microwave fields simultaneously.

The first of them is the electromagnetic pumping field with
frequency $F_{\mathrm{pump}} = F_{(0,-1)}$ and normalized amplitude
of magnetic component $Y$. This microwave magnetic component
interacts with the spin system, exciting $| \Delta M | = 1$
transitions between the lower $E_{(-1)}$ and the upper $E_{(0)}$
spin levels.

The second field is microwave ultrasound (called also hypersound)
with frequency $F_{\mathrm{signal}} = F_{(+1,-1)}$ and normalized
acoustic intensity $J$. This acoustic field interacts with the spin
system, exciting $| \Delta M | = 2$ transitions between the lower
$E_{(-1)}$ and the intermediate $E_{(+1)}$ spin levels. Such
unusual selection rule ($| \Delta M | = 2$) is caused by the
quadratic on $\widehat{\bm{\mathcal{S}}}$ Hamiltonian of
spin-phonon interaction
\cite{Altshuler-1972,Spin-phonon-ruby,Nickel-early,Tucker-etc} for
iron group ions in crystals.

Let us normalize the population differences $N_{(M,M')} \equiv
N_{(M)} - N_{(M')}$ at the spin transitions $E_{(M)} \rightarrow
E_{(M')}$ to their equilibrium values $N_{(M,M')}^{(e)} \equiv
N_{(M,M')} \bigl|_{ Y = 0, J = 0}$:
\begin{equation}
  {\mathsf{D}} \equiv  \frac{N_{(-1,0)}}{N_{(-1,0)}^{(e)}}; \quad
  {\mathsf{K}} \equiv - \frac{N_{(+1,-1)}}{N_{(+1,-1)}^{(e)}}, \label{eq:ap-C01}
\end{equation}
where $0 < {\mathsf{D}} < 1$ and $-1 \leq {\mathsf{K}} < L-1$ (note
sign ``minus'' in the definition of ${\mathsf{K}}$, Eq.
(\ref{eq:ap-C01})). Here ${\mathsf{K}}$ is the inversion ratio at
the $E_{(+1)} \rightarrow E_{(-1)}$ spin transition, and the
parameter $L$ characterizes properties of an active medium. Value
$L$ depends on the ratio of pump and signal frequencies
$F_{\mathrm{pump}} / F_{\mathrm{signal}}$, on the times of
longitudinal relaxation etc. The third population difference
$N_{(+1,0)}$ is univocally defined by ${\mathsf{D}}$ and
${\mathsf{K}}$, because $\sum_M \Bigl( N_{(M)} / N_{(M)}^{(e)}
\Bigr) = 1$.

For linear autonomous quantum amplifiers with weak signal ($J \ll
1$) standing-wave pumping becomes bistable
\cite{ZhTF-1991,ZhTF-1999} if parameter of cooperativity $C$ is
greater than its critical value $C_{\mathrm{cr}}$ (where
$C_{\mathrm{cr}}$ is codimension-2 bifurcation point), and if
$Y_{\downarrow} \leq Y \leq Y_{\uparrow}$ ($Y_{\downarrow}$ and
$Y_{\uparrow}$ are codimension-1 bifurcation points). For nonlinear
autonomous quantum amplifiers bifurcation values $C_{\mathrm{cr}}$,
$Y_{\downarrow}$ and $Y_{\uparrow}$ are renormalized by an intense
running-wave signal with constant amplitude
\cite{RiE-2003,Cen2ch5-2003}, and the collapse of inversion state
becomes possible.

Phaser amplifier with modulated microwave acoustic signal is
{\textit{nonautonomous}} (at least one CP in our dynamic system
becomes time dependent). But the equations of motions for
nonautonomous dynamic system may be transformed to equivalent
autonomous form in an extended phase space \cite{Parker-1987}. The
only requirement for this transformation is regularity of
modulation. In an extended phase space $({\mathsf{D}},
{\mathsf{K}}, \zeta) \subset \mathbb{R}^3$ (where $\zeta = \omega_m
{\mathsf{t}}$, $\omega_m$ is the modulation frequency, and
${\mathsf{t}}$ is usual, i.~e. non-discretized time) the equations
of motion for nonautonomous phaser amplifier with bistable pump
\cite{RiE-2003,Cen2ch5-2003} transforms to such the autonomous form
\cite{RiE-2004}:
\begin{equation}
  \begin{cases}
    {T_1 \partial \bm{\Delta}} / {\partial {\mathsf{t}}} = \bm{F} \bigl( \bm{\Delta}, \zeta / \omega_m, \bm{\Theta}_A \bigr); & \bm{\Delta}(0) = {\bm{\Delta}}_0;\\
    {\partial \zeta} / {\partial {\mathsf{t}}} = \omega_m;                   & \zeta (0) = \omega_m {\mathsf{t}}_0.
  \end{cases}
  \label{eq:ap-C02}
\end{equation}
Here $\bm{\Delta} \equiv ({\mathsf{D}},{\mathsf{K}})$;
$\bm{\Theta}_A$ is time-independent vector of CP; $\bm{F} \equiv
(F_{\mathsf{D}}, F_{\mathsf{K}})$. We suppose $T_1 \equiv T_1
\bigl( E_{(0)} \bigr) \approx T_1 \bigl( E_{(+1)} \bigr)$ (i.~e.
$\tau_e \approx \tau_r$) by arguments of Appendix \ref{ap:B}, and
the inequality characterizing class-B active systems $T_1 \gg
T_{\text{field}} \gg T_2$ takes place by definition
\cite{Tredicce-1985} (see also Section \ref{se:i} of the present
work). Components of vector $\bm{F}$ are as follows:
\begin{gather}
  F_{\mathsf{D}} =  1 - {\mathsf{D}} - \frac{Y^2 {\mathsf{D}}}{(1 + 2C{\mathsf{D}})^2} +
  \frac{\widetilde{J} {\mathsf{K}}}{4L};\label{eq:ap-C03}\\
  F_{\mathsf{K}} = -1 - {\mathsf{K}} + \frac{L Y^2 {\mathsf{D}}}{(1 + 2C{\mathsf{D}})^2} - \widetilde{J} {\mathsf{K}},\label{eq:ap-C04}
\end{gather}
where $\widetilde{J} = \bigl( 1 + k_{mJ} \sin\zeta \bigr) J_0$;
$k_{mJ}$ is the coefficient of modulation.

Eqs. (\ref{eq:ap-C02}) to (\ref{eq:ap-C04}) describe bistability,
collapse of inversion states, critical slowing-down and other
nonlinear processes in nonautonomous phaser amplifier (in framework
of lumped, i.e. point-like three-level model). An example of
critical slowing-down during collapse of inversion states is shown
at Figure~A3.

\section{LIST OF ABBREVIATIONS}\label{ap:D}

1D, 2D, ... --- One-Dimensional, Two-Dimensional, ...

1C, 2C, ... --- One-Channel, Two-Channel, ...

BZ --- Belousov-Zhabotinskii

BR --- Bogach-Reshodko

CA --- Cellular Automaton

CAU --- Cellular Automaton Unit

CP --- Control Parameters

CGL --- Conway's Game of Life

CML --- Coupled Map Lattice

GBC --- Geometry and Boundary Conditions

GH --- Greenberg-Hastings

GWR --- Generalized Wiener-Rosenblueth (model)

MSR --- Multi-Step Relaxation

OK2 --- Oono-Kohmoto 2D (model)

RtC --- Rays through Clouds (automaton)

RSW --- Rotating Spiral Wave

SSR --- Single-Step Relaxation

TLCA --- Three-Level Cellular Automaton

WR --- Wiener-Rosenblueth

ZM --- Zykov-Mikhailov

\clearpage

\begin{widetext}

\begin{center}

{\textbf{FIGURE CAPTIONS}}

to the paper of S.~D.~Makovetskiy and D.~N.~Makovetskii

``A Computational Study of Rotating Spiral Waves and
Spatio-Temporal Transient Chaos\\ in a Deterministic Three-Level
Active System''

(figures see as separate PNG-files)

\end{center}

Figure~1: {\textbf{Typical stages of TLCA evolution.}} CP set is as
follows: $\tau_e = \tau_r = h = 50$; $g = 1$; $f = 8$. Starting
pattern ${\mathrm{P}}^{(0)} = {\mathrm{P}}^{(0)}_A$ has $100 \times
100 = 10^4$ CAUs. Black, gray and white pixels denote excited,
refractory and ground-state CAUs respectively. Numbers under
patterns are equal to the steps $n$ of evolution. At $n = 3.66
\cdot 10^5$ the system is already at periodic attractor.

\vspace{10pt}

Figure~2: {\textbf{Divergence of spatio-temporal structures for}}
$\bm{f = 8}$ {\textbf{(upper row) and}} $\bm{f = 9}$
{\textbf{(lower row) during TLCA evolution.}} Values of $\tau_e,
\tau_r, h, g$ are the same as in Figure~1, but for another starting
pattern ${\mathrm{P}}^{(0)} = {\mathrm{P}}^{(0)}_C$ (see Figure~A1
for ${\mathrm{P}}^{(0)}_C$). At $n = 3 \cdot 10^5$ both systems are
at their attractors. Only excited CAUs are shown at this figure
(black pixels).

\vspace{10pt}

Figure~3: {\textbf{Slow evolution of TLCA and metamorphoses of RSW
topology.}} Starting pattern ${\mathrm{P}}^{(0)} =
{\mathrm{P}}^{(0)}_C$. Values  $\tau_e, \tau_r, h, g$ are the same
as in Figure~1 and Figure~2, but $f = 4$. Only excited CAUs are
shown at this figure (black pixels). The system reaches an
attractor with $Q_T = -3$ at $n \approx 4 \cdot 10^6$.

\vspace{10pt}

Figure~4: {\textbf{Collapse of excitations for TLCA by slow
evolution.}} Values $\tau_e, \tau_r, h, g, f$ are the same as in
Figure~3, but starting pattern is another: ${\mathrm{P}}^{(0)} =
{\mathrm{P}}^{(0)}_N$ (see the ${\mathrm{P}}^{(0)}_N$ at
Figure~A1). Transient time $n_C = 3385479$ is the lifetime of
nontrivial state of the whole cellular automaton with the pointed
starting pattern.

\vspace{10pt}

Figure~5: {\textbf{Developed transient spatio-temporal chaos in
TLCA.}} Set of CP is as follows: $\tau_e =9, \tau_r = 12, h = 25$;
$g = 1$; $f = 9$ (ZM model). Starting pattern ${\mathrm{P}}^{(0)} =
{\mathrm{P}}^{(0)}_C$. Lifetime $n_C = 2427063$. Black, gray and
white pixels denote excited, refractory and ground-state CAUs
respectively.

\vspace{10pt}

Figure~6: {\textbf{Pre-finish stage of transient process in TLCA
with \emph{f}~=~3. All RSWs have the same (by magnitude and sign)
effective topological charge}}. Black, gray and white pixels denote
excited, refractory and ground-state CAUs respectively. The perfect
left-handed multi-RSW pattern is shown for $n = 1.6 \cdot 10^5$.
The CP set is as follows: $\tau_e = \tau_r = h = 50$; $g = 1$; $f =
3$. This pattern is unstable. Multiple RSWs with $Q_T = - 1$
compete strongly after the left-handed pattern is formed, and "the
winner takes all" at $n > 3 \cdot 10^5$ --- the single RSW with the
same $Q_T = - 1$ occupies the whole active medium  (not shown). But
winner must die --- excitation collapses fully at $n_C = 396982$
due to collision of the winner's core with boundary (not shown).

\vspace{10pt}

Figure~7: {\textbf{Reversing of sign of effective topological
charge during TLCA evolution. Part~1: Almost purely right-handed
pattern.}} Step $n = 1.48 \cdot 10^6$ of evolution of starting
pattern ${\mathrm{P}}^{(0)} = {\mathrm{P}}^{(0)}_{BC}$ with $M_X =
M_Y = 900$ (pattern ${\mathrm{P}}^{(0)}_{BC}$ itself is not shown
here). The CP set is the same as at Figure~6. Only excited CAUs are
shown by black pixels. Multiple RSW with $Q_T = + 1$ forms
right-handed pattern. One may expect that winner will be
right-handed too. But the subsequent evolution of the system is
very surprising, see Figure~8.

\vspace{10pt}

Figure~8: {\textbf{Reversing of sign of effective topological
charge during TLCA evolution. Part~2: The single \emph{left-handed}
RSW is the result of evolution of the \emph{right-handed} pattern
of RSWs.}} This is one of the snapshots (namely at step $n = 2
\cdot 10^6$) of subsequent evolution of the pattern shown at
Figure~7. The winner RSW, shown here, has $Q_T = -1$, i.~e. the
system {\textit{becomes left-handed despite of their long
right-handed previous life}}. During further evolution ($n > 2
\cdot 10^6$, not shown here), the winner's core is moving
irregularly across the grid, it collides with the grid boundary,
and full collapse of excitations takes place at $n_C = 2260964$
with final stage by scenario shown at Figure~4.

\vspace{10pt}

Figure~9: {\textbf{Transient chimera states, i.~e. coexistence of
periodic and aperiodic spatio-temporal structures in TLCA
(demonstrated by sequences of generalized Poincar\'{e}
cross-sections).}} Starting pattern ${\mathrm{P}}^{(0)} =
{\mathrm{P}}^{(0)}_C$. Excited CAUs are black (TRUE), and
non-excited CAUs (both refractive and ground-state ones) are white
(FALSE). Parameters $\tau_e, \tau_r, h, g$ are the same as in
Figure~1, but $f = 5$. Central column of the Figure --- usual
sequence of TLCA patterns, left column --- forward-updated sequence
of ascending Poincar\'{e} cross-sections ${\mathcal{P}}^{(+)}(n_1,
\Delta n, k_p)$, right column --- backward-updated sequence of
descending Poincar\'{e} cross-sections ${\mathcal{P}}^{(-)}(n_2,
\Delta n, m_q)$. Here $n_1 = 154000$; $n_2 = 338000$;  $\Delta n =
1000$; $k_{\max} = m_{\max} = 184$ (i.~e. ending point for the
whole sequence of cross-sections ${\mathcal{P}}^{(+)}$ is the
starting point for ${\mathcal{P}}^{(-)}$ and vice versa).  The TLCA
at $n_2 = 338000$ is already at the attractor.

\vspace{10pt}

Figure~10: {\textbf{Symmetry restoring (left column) and
``breaking'' (right column) in TLCA.}} Black, gray and white cells
denote excited, refractory and ground-state CAUs respectively.
Dimensions of the patterns are $M_X = 75$; $M_Y = 50$ (grid lines
are shown for clarity). The CP set is as follows: $\tau_e = 5$;
$\tau_r = 7$; $g = 0.3$; $h = 3$; $f = 9$. Denotations of the rows
corresponds to the same steps of evolution for the left and the
right starting patterns: $a \rightarrow n = 0$;  $b \rightarrow n =
500$; $c \rightarrow n = 1256$; $d \rightarrow n = 10129$. Starting
patterns differ by the single CAU with $i = 66$; $j = 41$ (cells
are numbered beginning from the left bottom corner of the grid).
This CAU in the left starting pattern ${\mathrm{P}}_{ML}^{(0)}$ is
at level $L_{\mathrm{III}}$ (excited state, black cell). Analogous
CAU in the right starting pattern ${\mathrm{P}}_{MR}^{(0)}$ is at
level $L_{\mathrm{I}}$ (ground state, white cell).

\vspace{10pt}

Figure~11: {\textbf{Hypersensitivity to initial condition as cause
of unpredictability of left- or right-handed vorticity.}}
Dimensions of the patterns and the CP set are the same as at
Figure~10, but starting patterns ${\mathrm{P}}_{SL}^{(0)}$ (left
column, row $a$) and ${\mathrm{P}}_{SR}^{(0)}$ (right column, row
$a$) are slightly changed comparatively to
${\mathrm{P}}_{ML}^{(0)}$ and ${\mathrm{P}}_{ML}^{(0)}$.
Denotations of the rows corresponds to such the steps of evolution:
$a \rightarrow n = 0$; $b \rightarrow n = 500$; $c \rightarrow n =
2010$; $d \rightarrow n = 10118$. Starting patterns
${\mathrm{P}}_{SL}^{(0)}$ and ${\mathrm{P}}_{SR}^{(0)}$ differ by
the single CAU with the same coordinates $i = 66$; $j = 41$ as for
Figure~10. This CAU in the left starting pattern
${\mathrm{P}}_{SL}^{(0)}$ is at level $L_{\mathrm{III}}$ (excited
state, black cell). But analogous CAU in the right starting pattern
${\mathrm{P}}_{SR}^{(0)}$ is at level $L_{\mathrm{II}}$ (ground
state, gray cell).

\vspace{10pt}

Figure~A1: {\textbf{Samples of starting patterns.}} The population
of excited CAUs (relative quantity of black pixels) is equal to 0.4
for all these patterns. This is slightly less than population of
ground-state CAUs (white pixels), but more than twice greater than
population of refractory CAUs (gray pixels). Resulting inversion of
populations at $L_{\mathrm{III}} \leftrightarrow L_{\mathrm{II}}$
transition corresponds to an alternative variant of the phaser
medium activation scheme (in the original scheme of the
${\mathrm{Ni^{2+}:Al_2O_3}}$ phaser
\cite{Nickel-early,DNM-Diss-1983}, the analog of $L_{\mathrm{II}}
\leftrightarrow L_{\mathrm{I}}$ transition was inverted).

\vspace{10pt}

Figure~A2: {\textbf{Energy levels of divalent nickel in corundum.}}
This is the simplest system by which microwave phonon laser
(phaser) amplification, generation and inertial self-focusing were
realized at Gigahertz-range frequencies. At non-zero static
magnetic field, a collection of such active centers (interacting by
$d-d$ mechanism) is a kind of excitable three-level system with
two-channel diffusion of excitations. See Appendix \ref{ap:A} for
explanation of the energy levels structure and underlying
mechanisms of the levels splitting. For relaxation properties of
this system at low temperatures see Appendix \ref{ap:B}.

\vspace{10pt}

Figure~A3: {\textbf{Critical slowing down and collapse of inversion
states in a lumped three-level class-B active system.}}
Various-time oscillating routes to collapse are shown for the
inversion ratio ${\mathsf{K}}$, see Eqns. (\ref{eq:ap-C02}) to
(\ref{eq:ap-C04}) in Appendix \ref{ap:C}. The invariant subset of
CP (i.~e. unchanged components of $\bm{\Theta}_A$) is as follows:
$C = 5$; $L = 2$; $Y = 6$; $J_0 = 0.15$; $k_{mJ} = 0.7$. The
control parameter $\omega_m$ is varied relatively to relaxation
time $T_1$. Parameter ``${\mathsf{n}}$'' enumerates dependencies
${\mathsf{K}}({\mathsf{t}})$ for different values of $\omega_m
T_1$, namely: $({\mathsf{n}} = 1) \rightarrow (\omega_m T_1 =
0.1)$; $2 \rightarrow 0.15$; $3 \rightarrow 0.19$; $4 \rightarrow
0.199$; $5 \rightarrow 0.2$; $6 \rightarrow 0.2004$; $7 \rightarrow
0.2005$; $8 \rightarrow 0.20053$. Initial conditions:
${\mathsf{D}}({\mathsf{t}}=0) = 0.01$; $\mathsf{K}({\mathsf{t}}=0)
= 0.7$; $\zeta({\mathsf{t}}=0) = 0$.

\clearpage

\end{widetext}

\end{document}